\documentclass[titlepage, 11pt]{osumkt-article}

\let\svthefootnote\thefootnote
\newcommand\blankfootnote[1]{%
  \let\thefootnote\relax\footnotetext{#1}%
  \let\thefootnote\svthefootnote%
}

\title{Advertising and Brand Attitudes: Evidence from 575 Brands over Five Years
\blankfootnote{\hspace{-1.8em}The authors thank Wes Hartmann, two anonymous reviewers, and numerous seminar audiences for helpful comments and discussions. This study was made possible by the authors' employers, and data were drawn from standard Kantar and YouGov data sources, but the analysis is the authors' alone as it was not funded or otherwise influenced by any other party.}
}
\author{Rex Yuxing Du \\ \small{Bauer Professor of Marketing} \\ \small{Bauer College of Business} \\ \small{University of Houston} \\ \small{rexdu@bauer.uh.edu} \\ \\ Mingyu Joo \\ \small{Assistant Professor of Marketing} \\ \small{School of Business} \\ \small{University of California, Riverside} \\ \small{mingyu.joo@ucr.edu} \\ \\ Kenneth C. Wilbur \\ \small{Associate Professor of Business Analytics and Quantitative Marketing} \\ \small{Rady School of Management} \\ \small{University of California, San Diego} \\ \small{kennethcwilbur@gmail.com}}
\date{\today}

\usepackage{natbib}
 \bibpunct[, ]{(}{)}{,}{a}{}{,}%
\usepackage{color,soul}

\usepackage{xcolor,colortbl}

\usepackage{pdflscape}
\usepackage{arydshln} % for Dashed Line
\usepackage{booktabs} % for much better looking tables
\usepackage{array} % for better arrays (eg matrices) in maths
\usepackage{paralist} % very flexible & customisable lists (eg. enumerate/itemize, etc.)
\usepackage{verbatim} % adds environment for commenting out blocks of text & for better verbatim
\usepackage{subfig} % make it possible to include more than one captioned figure/table in a single float
\usepackage{amsmath}
\usepackage{ulem}

\makeatletter
\newcommand{\leqnomode}{\tagsleft@true}
\newcommand{\reqnomode}{\tagsleft@false}
\makeatother

\usepackage{amssymb}
\usepackage{setspace}
\usepackage{lscape}
\usepackage{graphicx}

\usepackage{enumerate}

\renewcommand{\emph}{\textit}

\begin{document}
\maketitle
\newpage

\begin{center}
\renewcommand{\baselinestretch}{1}
\Large{Advertising and Brand Attitudes: Evidence from 575 Brands over Five Years}
\end{center}
\vspace{1em}
\renewcommand{\baselinestretch}{1.4}
\begin{center}
{\textbf{Abstract}}
\end{center}

\noindent Little is known about how different types of advertising affect brand attitudes. We investigate the relationships between three brand attitude variables (perceived quality, perceived value and recent satisfaction) and three types of advertising (national traditional, local traditional and digital). The data represent ten million brand attitude surveys and \$264 billion spent on ads by 575 regular advertisers over a five-year period, approximately 37\% of all ad spend measured between 2008 and 2012. Inclusion of brand/quarter fixed effects and industry/week fixed effects brings parameter estimates closer to expectations without major reductions in estimation precision. The findings indicate that ({\it i}) national traditional ads increase perceived quality, perceived value, and recent satisfaction; ({\it ii}) local traditional ads increase perceived quality and perceived value; ({\it iii}) digital ads increase perceived value; and ({\it iv}) competitor ad effects are generally negative. %Comparisons with descriptive models present that inclusion of control variables for unobserved confounds leads to parameter estimates closer to expectations without major reductions in estimation precision.
%The results ...call into question? ...show that, while descriptive models present strong associations between advertising and brand attitudes, models with extensive time-varying control variables present limited effects of advertising on brand attitudes. The comparison between two different identification strategies suggests that the extant descriptive media mix models may mislead firms to be overconfident in their advertising decisions.
\vspace{1em}
\begin{flushleft}
\end{flushleft}

\newpage
\setlength\arraycolsep{2pt}
%\layout

\section{Introduction}%\footnote{\textcolor{gray}{\it Try to say: `recent satisfaction' (not `satisfaction'); `brand attitudes' not `brand tracking;' `industry' not `category;' `advertising expenditure' or `ad spend' , broken into `national traditional,' `local traditional' and `digital'}}
Advertising practitioners describe two types of advertising goals: direct response and brand attitudes.\footnote{The same labels are also applied to advertising {\it content}, which typically reflects the goals of the ad campaign, but is regrettably unobserved in our dataset.} Direct response goals incorporate short-run reactions to ads, such as phone calls, store visits, website traffic or online sales. Brand attitude goals incorporate long-run reactions to ads, such as consumers' perceptions of quality, value or satisfaction. The two goal types often overlap as both seek to influence bottom-line objectives (e.g., total sales, profits), but the difference in time horizon makes direct response goals more easily attributable to ads than brand attitudes. The difficulty of brand attitude attribution leads many advertisers to forego attribution; only 50\% of Chief Marketing Officers say they prove the short-term impact of marketing spend on the business quantitatively, and just 41\% say they prove the long-term impact quantitatively.\footnote{https://cmosurvey.org/wp-content/uploads/sites/15/2018/02/The\_CMO\_Survey-Topline\_Report-Feb-2018-1.pdf, accessed March 2018.}
%Executives believe that direct response ads dominate online advertising, whereas brand advertising predominates in traditional media.\footnote{https://www.comscore.com/ita/Insights/Blog/On-Branding-Versus-Direct-Response-Advertising, accessed August 2018.} 
% [Ken: The sentence of executives' categorization of online as short-term vs. traditional as long-term may be inconsistent with brand-lift in Facebook and Google, and it may limited interpretation of our results or may require further thoughts or investigation of our advertising data to interpret the results.]

%\textcolor{blue}{Both types of advertising goals seek to ultimately influence bottom-line objectives like sales and profits, and they may frequently overlap. However they differ in time horizon and measurability. In particular, long-term advertising goals may be incompatible with common approaches to measuring immediate responses to advertising campaigns.\footnote{As an example, 50\% of Chief Marketing Officers say they prove the short-term impact of marketing spend on the business quantitatively, and 41\% say they prove the long-term impact quantitatively (https://cmosurvey.org/wp-content/uploads/sites/15/2018/02/The\_CMO\_Survey-Topline\_Report-Feb-2018-1.pdf, accessed March 2018.)}}

There is some controversy in the academic literature about marketers' ability to estimate the effect of ad spend on bottom-line goals such as sales and revenue. For example, although TV advertising experiments are scarce\footnote{Advertising experiments are scarce in general; see, e.g., Rao and Simonov (\citeyear{rs:2018}).} and some have exhibited limited statistical power in split-cable designs (e.g., \citealt{Letal:1995}), quasi-experimental research has estimated precise effects of TV ads on direct response goals (e.g., Tellis et al. \citeyear{TCT:2000}, Liaukonyte et al. \citeyear{LTW:2015}, Du et al. \citeyear{DWX:2017}, Hartmann and Klapper \citeyear{HK:2018}, Shapiro \citeyear{S:2018}). On the other hand, recent display advertising field experiments have shown convincingly that extremely large sample sizes are required to adequately power advertising experiments and that observational methods may be poor substitutes for experimental estimates (Lewis and Rao \citeyear{LR:2015}, Gordon et al. \citeyear{gzbc:2017}).\footnote{Digital advertising delivery facilitates experimentation and the measurement of individual-level response data, but the advertising medium is beset by several widespread problems that complicate experimental analysis, including ad (non-)viewability (IAB \citeyear{IAB:2015}), a high incidence of ad blocking by default (Shiller et al. \citeyear{SWR:2018}), non-human traffic (WhiteOps \citeyear{W:2016}), and advertising blindness (e.g., Owens et al. \citeyear{OPC:2014}). It remains unclear whether such display advertising results apply to other media.}

The purpose of the current paper, broadly, is to investigate the links between brand attitude data and advertising expenditures in a large sample of brands that advertise regularly. Although sales data are sparse and highly variable, brand attitude data tend to be non-sparse and highly stable. Many large brands have subscribed to ``brand tracking'' surveys for decades, and the supply of such data may be increasing. For example, Facebook and Google both recently introduced products to estimate ``lifts'' in brand attitudes resulting from advertising.\footnote{Note that we do not have data on direct response goals and therefore cannot speak to such results.}

Although sales data are usually the most important indicator of advertising effects, they may not be the only, or even the best, statistical indicator of advertising response for all brands. Some firms -- particularly those whose products exhibit long purchase cycles or long inter-purchase times -- may prefer to consider intermediate response variables such as brand attitudes. Consumers' brand attitudes are important indicators in their own right, as they reflect consumer perceptions about brand quality and value, and predict downstream behaviors such as search and consideration (\citealt{DFFOY:2017}). The financial value of brand attitudes are made tangible by brand asset valuations; strong brands often sell for substantially more than physical asset valuations because consumer attitudes tend to persist, even when a brand changes owners. From a practical perspective, many advertisers cannot estimate causal effects of ads on sales, yet they still face operational questions such as whether to advertise; how much to spend; and how to allocate their expenditure across media. One possible way forward is to consider replacing sales data with other measures of consumer response to advertising.

More specifically, we address three primary questions. How do brand attitudes change with advertising by the same brand and its competitors? How do these relationships vary across attitude measures and types of advertising media? How do various strategies to control for time-varying unobservables change effect sizes and precision? Our goal, to the extent possible, is to ``let the data speak'' by applying comparable methods to comparable measures for many advertising brands. %We also investigate how effect sizes vary across industries, and whether changes in brand attitudes predict advertising expenditures.{%3) How much do the differences between descriptive and causal estimates vary across media types?

To answer these questions, we examine a unique dataset of 575 established brands from 37 industries over a five-year observation window, merging weekly brand attitude data with weekly advertising expenditure data. In totality, the data include \$264 billion spent on advertising, 37\% of all ad spend measured during the observation window, and approximately ten million brand attitude surveys. We study mature brands that advertise regularly in a ``large-N, large-T'' panel dataset. 
%have a similar scope to meta-analytic approaches (e.g., \citealt{STB:2011}).

The brand attitude metrics we consider are the percentages of survey respondents indicating favorable perceived quality, perceived value and recent satisfaction for each brand in each week. The three types of advertising media we consider are national traditional media, local traditional media, and digital media. We suspect that each type of advertising could operate directly on each brand attitude: advertising content may communicate differentiating features, thereby influencing perceived quality; it could communicate current pricing terms, thereby influencing perceived value; and it could lead a consumer directly to purchase, thereby increasing the proportion of people who indicate recent satisfaction. Although we believe that any of these effects {\it may} operate, we expect perceived quality to be most strongly linked to national advertising, as national ads typically convey product information and differentiating messages (\citealt{LTW:2015}). We expect perceived value to rise more strongly with local traditional advertising and digital advertising, as pricing and availability frequently vary geographically and such information is often communicated via advertising in geographically targeted media (Kaul and Wittink \citeyear{KW:1995}, Lee et al. \citeyear{LHN:2017}, Xu et al. \citeyear{XWSS:2014}). 

The models we estimate all include lagged brand attitudes, contemporaneous and lagged ad spending by type of media, brand fixed effects, week fixed effects, and weighted standard errors to reflect exogenous variation in the number of survey respondents each week. The fundamental challenge to causal inference in this setting is not in the nonrandom assignment of advertisements to consumers; brand attitude data are collected from large samples of consumers whose selection is plausibly unrelated to advertising efforts. Instead, there is a primary difficulty in the {\it timing} of advertising expenditure, as advertising timing may be nonrandomly selected and could coincide with periods of peak demand or heightened responsiveness to advertising.

We investigate two sets of control variables as possible remedies to this advertising timing endogeneity problem: brand/quarter fixed effects, to control for time-varying, brand-related unobservables that may drive both advertising and brand attitudes; and industry/week fixed effects, to control for industry-level unobservables that may affect multiple competing brands' advertising and brand attitudes. When both sets of control variables are included, causal interpretation requires an assumption that brand/week advertising expenditures are not chosen with knowledge of future brand/week departures from brand/quarter unobservables or future brand/week departures from industry/week unobservables. Although this identifying assumption is unlikely to apply to every brand, we suspect it applies to the large majority of brands in the sample. %The results indicate that inclusion of both sets of control variables leads to advertising estimates that are more intuitive without much loss of precision. 

To summarize the primary findings, the data indicate that brand/quarter and industry/week fixed effects are individually and jointly important determinants of brand attitude data. Further, the model that includes both sets of control variables produces results that comport better with expectations, and exhibit greater internal coherence, than a descriptive model without either set of control variables. The estimates indicate that ({\it i}) brand attitude metrics all rise with multiple lags of the brand's own national traditional advertising; ({\it ii}) local traditional ads increase perceived quality and perceived value; ({\it iii}) digital ads increase perceived value; ({\it iv}) competitors' ads reduce brand attitudes. 

The results come with two important caveats. First, we are not able to observe direct response outcomes for such a large and diverse sample of brands. In cases that we do not find a significant effect of an advertising medium on a brand attitude, it does not prove that effect is zero, and it also says nothing about the effect of the advertising medium on direct response goals. Second, we interpret the estimates as Average Treatment Effects (ATE), subject to some survey sampling caveats discussed further below. However, we suspect that advertisers may care most about Treatment-on-the-Treated (ToT) effects, which will normally be larger.

Next, we discuss how the current study relates to extant literature. The subsequent sections explain the data and provide some model-free evidence; discuss identification; specify the empirical models; report and interpret the findings; and discuss the overall learnings, limitations and implications of the exercise.

\subsection{Relationship to Previous Literature}

%The importance of brand attitudes as intermediate measures of advertising responsiveness was first proposed by the experimental literature. Numerous experimental studies have shown that consumers' preferences are formulated by their affective responses, which are influenced by exposure to advertising (e.g., \citealt{BR:1986}). Therefore, creating a positive association between a brand and its image became one of advertising's popular goals \citep{AH:1987}. By the nature of the studies, the literature clearly shows the direction and causality of the relationship, but does not offer a reference of the effect sizes in marketplaces.

The empirical literature on advertising is vast. Most relevant is the set of papers that demonstrates that advertising can affect intermediate consumer outcomes, i.e., behaviors and attitudes that occur prior to sales. For example, \cite{dk:2011} show that advertising increases brand awareness and expands consumer choice sets; \cite{jwcz:2014} found that TV advertising increases the number of product category-related Google searches and the proportion of searches that contain brand-specific keywords; and \cite{HDD:2014} show that advertising predicts monthly search for automotive brands, which in turn predicts monthly purchase data. There are also several papers that estimate industry-specific relationships between brand attitudes and advertising expenditures (Hanssens et al. \citeyear{HPSVY:2014}, Srinivasan et al. \citeyear{SVP:2010}).

The most closely related paper is \cite*{CDD:2009}, which estimated advertising effects on brand awareness and perceived quality in a large annual panel dataset, including \$96B in ad spending by 348 brands from 2000\--2005. As that paper explains, most of the prior literature was based on cross-sectional data, with questionable ability to separate effects of advertising from unobserved confounds such as product quality. \cite{CDD:2009} found, in their preferred specification, that a focal brand's own advertising increased its own awareness but did not significantly change perceived quality. The focal brand's competitor advertising, by contrast, reduced brand awareness and increased perceived quality. Although the current analysis replicates some aspects of \cite{CDD:2009}, our incremental contribution rests on several important differences: temporal disaggregation, methods, measures and results. 

The most important difference may be the temporal dimension of the data. \cite{CDD:2009} analyzed a ``large-N, small-T'' type panel with 4.2 observations available for the average brand. The current paper, by contrast, investigates a balanced panel of 575 brands over 252 weeks of data, consistent with the central findings of \cite{TF:2006} that ``too disaggregate data does not cause any disaggregation bias.'' More granular data allow for more extensive controls for possible time-varying confounds, one of the central themes of our paper. In fact, Clark et al. (\citeyear{CDD:2009}, p. 229) said ``Perhaps the ideal data for analyzing the effect of advertising are time series of advertising expenditures, brand awareness, and perceived quality for the brands being studied. With long enough time series we could then try to identify for each brand in isolation the effect of advertising expenditures on brand awareness and perceived quality.'' %\footnote{\cite{CDD:2009} say this in the context of explaining why they do not estimate brand-specific advertising effects on brand attitudes. Although heterogeneous effects are interesting, the current paper focuses mostly on the overall average effects of advertising, in large part because brand-specific effects are not well identified by the data.} 
Intuitively, the more disaggregated data allows for a sharper delineation of the lead/lag relationships between the timing of ad spend (which is highly variable over time) and brand attitudes (which mostly exhibit stable long-run averages). There is further interest in contrasting results based on their 2000-2005 sample period with the later time period of 2008-2012, as consumer media habits and firm ad spending changed significantly between these two time periods; for example, digital advertising increased substantially.

There are also important differences in attitude measures, methods and results. \cite{CDD:2009} observed average ratings of perceived quality on a 0-10 scale, and defined awareness as the percentage of respondents who rated the brand's quality. The metrics studied in this paper indicate multiple dimensions of brand attitudes, including one (recent satisfaction) which may reflect recent purchase activity; but they do not explicitly separate awareness from other attitudes. We further distinguish between the effects of three types of ad spend (national traditional, local traditional and digital). \cite{CDD:2009} relied on dynamic panel instrumental variables estimators to control for advertising endogeneity, with findings that differed qualitatively across estimators. The exogeneity conditions require knowledge about the serial correlation of the error terms, information which is difficult to derive from theory or test in ``small-T'' settings. Finally, the empirical findings differ substantially: we find positive effects of own ad spend on perceived quality; we offer the first findings related to perceived value, recent satisfaction and individual types of advertising; and we find that competitor ad spending generally decreases brand attitudes.

The current study also relates to published meta-analyses of advertising effects (e.g., \citealt{STB:2011}). However, because the brand sample was selected systematically from a nearly comprehensive set of large advertisers, it may include more null effects than any given set of published case studies, suggesting the mean effect estimates may be more conservative and more representative.%whereas some existing meta analyses of published case studies  may omit brands or industries with null advertising effects. Inclusion of such brands and categories may have led to lower average effect sizes in the current study. Second, our survey sample selection is independent of ad targeting and response measures are adjusted to demographically represent the national population. Therefore, our estimates offer empirical lower-bounds of average treatment effects on representative individuals or on treated individuals with random selection.}%\footnote{\textcolor{blue}{The survey sample potentially under-represents treated individuals with narrow or specific targeting as in digital ads (e.g., \citealt{BNT:2015,LR:2015}).}}}

The current study is further related to a set of papers comparing advertising effects across media and across competitors. For example, \cite{DD:2013} offered an approach to help brands evaluate relative media effectiveness by linking loyalty program members' purchases to their responses on a media consumption survey. \cite{DHS:2014} showed that television advertisements produced statistically indistinguishable ``lift'' in aided brand recall to three formats of online advertisements (video, banner and rich media); but proper inference depends critically on accounting for differences in pre-existing brand knowledge between people exposed to different ad formats. \cite{lpx:2017} investigated a large panel of brands, showing that internet and television ad spend both have small but significant positive effects on word-of-mouth. There is also evidence that competitor advertising can interfere with advertisement recall \citep{KA:1994} and sales response \citep{DBD:2008}.

More broadly, the current study relates to the literature that estimates advertising effects on brand equity. \cite{ALN:2003} introduced estimation of customer-based brand equity and reported a positive association between advertising and brand equity. \cite{BGHM:2017} found that advertising investment increases the expected net present value of future cash flows due to a brand in a dynamic model of advertising investment. \cite{MGL:1997} found that advertising makes consumers less price sensitive and reduces the size of the non-loyal segment. Our results offer evidence consistent with possible attitude-related mechanisms underlying these important findings.

\section{Data and Model-free Evidence}
Two large-scale commercial databases are combined -- brand attitude survey data from YouGov and ad spending data from Kantar. We believe both data sources to be ``best in class.'' Both Kantar and YouGov are leading market research agencies.\footnote{https://www.ama.org/publications/MarketingNews/Documents/2017-top-50-gold-report-article.pdf, accessed March 2018.} To the best of our knowledge, there are no data sources that provide both better quality and similar coverage. We further believe that these two databases are the market leaders in their product categories, suggesting that we are using similar data to what many practitioners have available. However, the data do have some nuances that are important to consider when interpreting the results of the analysis. We first describe the data sources and focal metrics, then the sample selection, followed by summary statistics and model-free evidence.

\subsection{Brand Tracking Data}
Brands employ market research firms to conduct longitudinal surveys to monitor consumers' brand attitudes. Although such surveys have traditionally been quite costly, there are numerous research agencies that offer similar products, including GfK, Millward Brown, TNS and YouGov. Recently, Facebook introduced its own survey platform to enable brands to ``accurately measure brand awareness, impact and ad recall.''\footnote{https://www.facebook.com/business/learn/facebook-brand-polling, accessed March 2018.} The weaknesses of survey data are numerous and well documented. However, regular surveys of large consumer panels produce brand attitude data that are reasonably stable over time, although individual data points can be affected by sampling error. When meaningful changes do occur, they often correspond to identifiable shocks, such as news events or quality changes. %Model-free evidence in Figure \ref{fig:Evid}, below, provides support for these claims.

%These surveys provide reliable measurements of brand attitudes from large samples that are affordable to collect and readily comparable across brands, industries and time periods. There are dozens of marketing research firms offering syndicated brand tracking surveys.

Brand tracking data were drawn from the largest available survey panel, the YouGov BrandIndex. YouGov maintains a panel of more than 1.5 million U.S. consumers, with each panelist invited to complete up to one survey online each month. Panelists are compensated with redeemable ``points" each time they complete a survey, but survey participation is not mandatory, leading to some exogenous fluctuations in the number of surveys completed for each industry in each week.

\begin{figure}
\caption{Survey Instrument Example}
\label{fig:YouGovScrn}
\begin{center}
\vspace{-0.2cm}
\includegraphics[scale=0.7]{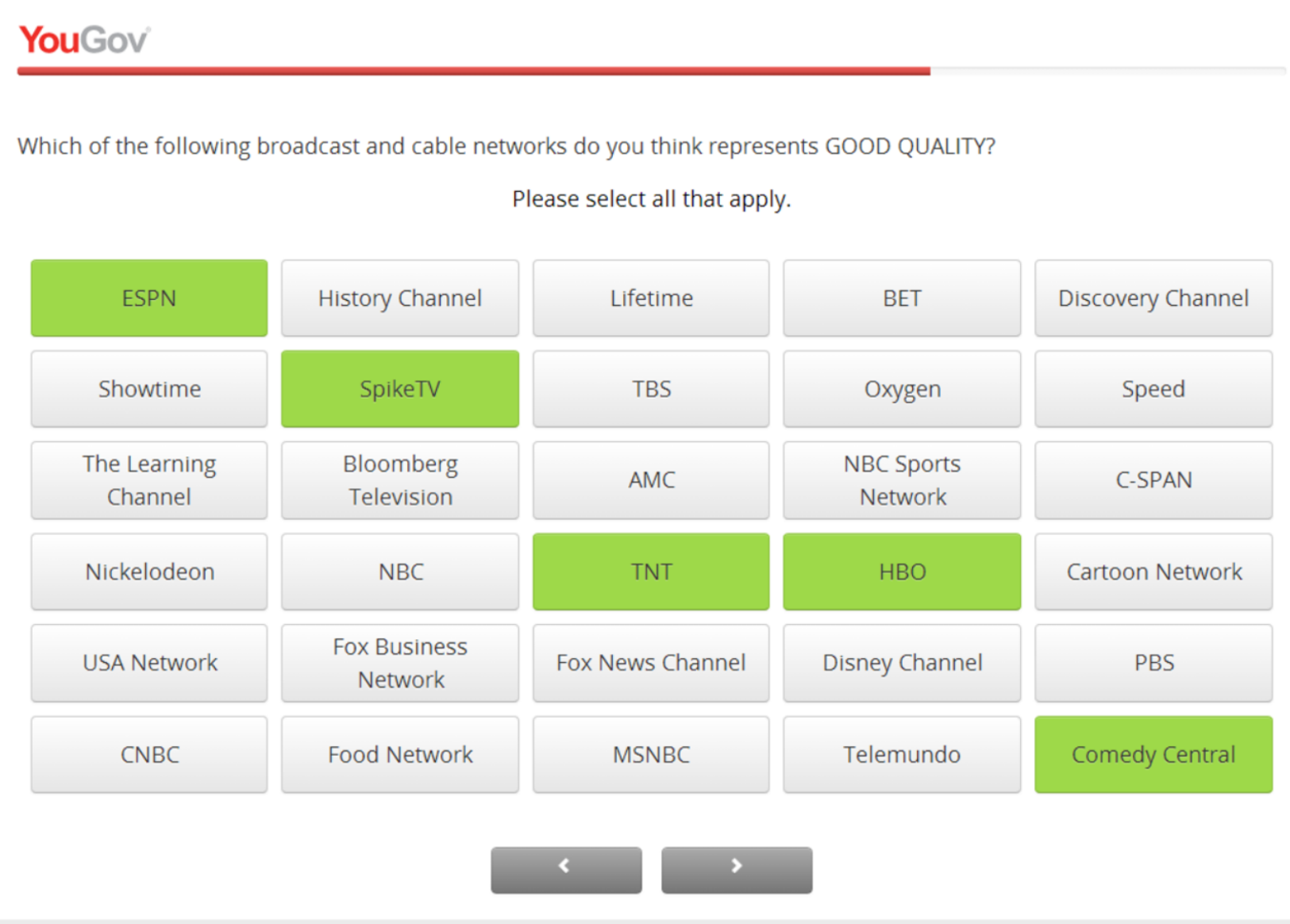}
\vspace{-0.4cm}
\end{center}
\end{figure}

Each survey respondent was asked one of seven attitude questions about seven different industries, with a different question for each industry. %\footnote{Asking each question about a different industry helps to prevent ``halo effects," i.e., a respondent tendency to mark the same set of brands in response to multiple different imagery questions.} 
The standardized response format, depicted in Figure \ref{fig:YouGovScrn}, solicited responses for 25-40 brands within each industry. %\footnote{Upon inspection, it appears to us that the set of brands in each industry was coherent. We believe YouGov based the industry classifications on feedback from client brands.} 
The survey instrument asked, for example, ``Which of the following broadcast and cable networks do you think represents good quality?" and then lists thirty television networks in random order. The respondent could mark as many brands as desired with no time limit, suggesting that the data should reflect absolute levels of quality, as perceived by the respondent.

YouGov collected data using the following set of questions:
\begin{itemize} \itemsep -2pt
\item ``Which of the brands do you associate with good quality?"
\item ``Which of the brands do you associate with good value-for-money?"
\item ``Would you identify yourself as a recent satisfied customer of any of these brands?"
\item ``Which brands would you recommend to a friend?"
\item ``For which brands do you have a `generally positive' feeling?"
\item ``Which of the brands would you be proud to work for?"
\item ``Over the past two weeks, which of the following brands have you heard something positive about (whether in the news, through advertising, or talking to friends and family)?"
\end{itemize}
The survey items remained constant throughout the sample period.

YouGov uses respondent demographics to weight the data and construct nationally representative averages, so the brand attitude data indicate the weekly percentage of U.S. consumers that would provide a positive response to each of these seven questions for each brand, and further indicate the weekly number of respondents answering each question for each industry. An appealing feature of this survey panel is that its selection is seemingly unrelated to advertising treatment. However, as with other incentivized-participation or permission-based survey samples, we are unable to rule out the possibility that survey respondents may have been nonrandomly selected on unobserved attributes, e.g., media consumption or proclivity to notice brands that advertise to them.

An important limitation of the brand attitude data is that some level of awareness is presumably required to provide a positive indication for a brand. For example, in Figure \ref{fig:YouGovScrn}, a respondent who has never heard of the Speed network will presumably not indicate that the network represents good quality. We view this as a regrettable but reasonable limitation of the brand attitude data. Some level of consumer awareness or familiarity is a prerequisite to the brand attitudes that we are able to observe.

We focus our study on the metrics of perceived quality, perceived value and recent satisfaction. Perceived quality and perceived value both relate to identifiable messages that are frequently communicated through advertising, such as differentiating statements about product attributes or current pricing terms. Recent satisfaction is the brand attitude metric that comes closest to indicating sales; if advertising increases sales, then it should also lift the proprotion of consumers who indicate that they are recently satisfied customers of the brand. %We label the percentage of consumers reporting positive perceptions of brand quality, value and recent satisfaction for brand $b$ in week $t$ as \textcolor{red}{$qua_{bt}$, $val_{bt}$ and $sat_{bt}$} respectively; and the numbers of respondents in each survey \textcolor{red}{for industry $i$ as $n_{it}^q$, $n_{it}^v$ and $n_{it}^s$,} respectively.

\subsection{Advertising Expenditure Data}
Kantar Media compiles comprehensive data on advertising placements and expenditure estimates across the broad range of advertising media listed in Table \ref{tb:mediacat}. Kantar is widely viewed as the market leader in ``competitive advertising intelligence,'' i.e., the service of monitoring competitors' advertisement placements and expenditures. 

Kantar tracks television, print and digital media by logging brand advertisement insertions algorithmically through continuous monitoring of media content. For television and print media, estimated advertising prices are provided by media outlets indirectly through the Standard Rate and Data Service (SRDS). Although the SRDS price estimates are known to be imperfect, they are commonly used by brands to plan future advertising efforts, and are the only available source of widespread information about advertising prices.\footnote{The reporting incentives are mixed. A media outlet could exaggerate its ad price to offer perceived discounts in negotiations with advertisers. Or, a media outlet might underreport its ad price to attract interested advertisers. Actual ad prices in traditional media are typically set in confidential bilateral negotiations and may reflect price discrimination or quantity discounts. Digital advertising prices are typically set in complex, rapidly changing spot auction markets within or between ad networks, demand-side platforms and supply-side platforms.} Digital advertising placements are collected by an elaborate system of web crawlers. Outdoor and radio ad placements and prices, and digital advertising price data, are provided directly to Kantar by industry partners.

We paid particular attention to Kantar's data quality in internet display and internet search data, as these measures were relatively new at the time we collected the data. The internet search ad spend data did not appear reliable: they were unreasonably sparse. Our investigations of internet display data did not indicate any identifiable problems. Therefore our measure of digital advertising includes internet display media only. We remain cognizant of the possibility of classical errors-in-variables problems which may bias parameter estimates toward zero and bias t-statistics downward (Griliches \citeyear{g:1977}), thereby yielding false null results. However, we do find significant effects of own digital and competitor digital advertising on brand attitudes.

%The advertising placement and expenditure data are obtained from Kantar Media's Stradegy database. This database tracks advertising insertions in national, local, and digital advertising media including 18 television, print, radio, outdoor, internet display and internet search media, as listed in Table .  Kantar is widely considered to be the leading provider of competitive advertising intelligence.

\renewcommand{\arraystretch}{1.1}
\begin{table}%[tb]
\caption{Advertising Media Tracked in Stradegy Database}
%\vspace{1em}
\label{tb:mediacat}
\begin{center}
{
\begin{tabular}{l|l|l}
\hline
\textbf{National Media} & \textbf{Local Media} & \textbf{Digital Media} \\
\hline
Business-to-Business & Local Magazines & Internet Display \\
Cable TV & Local Newspapers & Internet Search \\
National Newspapers & Hispanic Newspapers & \\
Magazines & Sunday Magazines & \\
Hispanic Magazines & Spot TV & \\
National Spot Radio & Syndicated TV & \\
Network Radio & Local Radio & \\
Network TV & Outdoor & \\
Spanish Language TV & & \\
\hline
\end{tabular}
}
\end{center}
\end{table}

Advertising content varies by type of media. National traditional ads are often used to communicate information and differentiating messages \citep{LTW:2015}, while local traditional ads focus more on current price and availability, as these variables typically vary across local markets, while also conveying some quality-relevant information \citep{KW:1995}. Digital advertising also frequently communicates current pricing and availability. For example, \cite{LHN:2017} quantified the contents of 100,000 Facebook ads. They reported that 62\% of ads offered deals (``discounts or freebies''), 44\% compared prices, and 69\% contained information on where to obtain a product. \cite{YLTP:2015} quantified search advertisement content for hotels, travel intermediaries and auto manufacturers, finding that they used 14\%, 25\% and 6\% of space, respectively, to communicate pricing terms.%\footnote{To the extent that price advertising is paid for by local retailers or distributors \citep{XWSS:2014}, it will not appear in our dataset.} 

Local traditional ads offer better targeting than national traditional ads, as they may vary across geographic markets. Digital ads can be even better targeted, based on demographic and behavioral variables, as well as geographically. A few recent studies have found that banner advertisements can increase sales (e.g., \citealt{LR:2014}), internet video ads have been found to be as effective as TV ads in brand building (\citealt*{DHS:2014}), and digital advertising revenues have grown much faster than traditional advertising in recent years. On the other hand, some published research has called digital ad effectiveness into serious question (\citealt{BNT:2015}, \citealt{LR:2015}); digital ads are subject to higher levels of ad non-viewability, passive ad blocking, ad blindness and non-human traffic; and some prominent brands including GM and P\&G have publicly questioned whether digital campaigns are cost-effective.\footnote{https://www.wsj.com/articles/SB10001424052702304192704577406394017764460, https://www.wsj.com/articles/p-g-cuts-more-than-100-million-in-largely-ineffective-digital-ads-1501191104, accessed March 2018.} 

The differences in advertising content and targeting across categories of advertising media lead us to suspect that relationships between brand attitudes and advertising expenditures may vary across these three categories of advertising media. %We denote the advertising expenditure of brand $b$ in week $t$ in national media, local media and digital media as $n_{bt}$, $l_{bt}$, and $d_{bt}$, respectively.

%National ad spending of brand $b$ in week $t$ ($n_{bt}$) is defined as weekly ad spending by the brand placed in national media in the first column of Table \ref{tb:mediacat}. National ads are expected to influence consumers' perception of quality but not perceived value for money, due to the nature of contents. Local ad spending of brand $b$ in week $t$ ($l_{bt}$) is weekly ad spending in local media in the second column of Table \ref{tb:mediacat}, and is expected to mainly influence perceived value of the brand. Digital ad spending ($d_{bt}$) is weekly ad spending in online search and display ads. We do not have clear expectation of consequences of digital ads on quality or value perception, as many digital ad campaigns are directly related to the last step of purchase funnel. However, some spillover effects of digital ads on traditional domain have been observed (e.g., \cite{LR:2014}), and internet ads are found to be as effective as TV ads in brand building \citep{DHS:2014}. Therefore, it may be important to understand how digital ads influence brand attitude metrics. We do not have clear expectation about ad effectiveness on satisfaction, because it may mostly be driven by past usage of services of products at least for experience goods.

\subsection{Brand Sample Selection}

The goal of this study is to estimate relationships between advertising expenditures and brand attitudes for mature brands that advertise regularly. We select brands with these particular criteria in mind, so we begin with the acknowledgement that the results can only be interpreted as applicable to the set of brands studied and may not generalize beyond that set. %\footnote{New brands or irregularly-advertising brands likely require more customized data and modeling techniques, frustrating the goal of generating comparable parameter estimates across brands and industries by holding the analysis constant.} 
Although this strategy does not represent the full population of brands, this subset is large and particularly important, as it accounted for 37\% of all advertising expenditure measured during the sample period. 

We first matched each brand in the YouGov data to its equivalent entity in the Kantar database. We then downloaded weekly ad spend data for each brand in each equivalent time period. Finally, we retained brands that (a) were tracked by YouGov for the entire sample period, (b) advertised in at least 30\% of the observed weeks, and (c) did not go more than thirteen consecutive weeks with zero advertising. The set of 575 brands meeting these criteria is provided in the appendix, along with each brand's industry as indicated by YouGov. In total, these brands spent \$264 billion on advertising from 2008-2012, or \$92 million per brand per year. The corresponding brand attitude metrics are based on about ten million surveys, yielding a weekly average of 595 responses per question per brand (SD=77).

\subsection{Descriptive Statistics and Model-free Evidence}

We first summarize the brand attitude data, followed by the ad spend data. We then visualize the relationships between them for a few selected brands, then present quantiles of brand-level correlations between the key variables in the analysis.

\begin{figure}[htp]
\caption{Brand Attitudes by Industry and Year}
\label{fig:MeanBA}
\begin{center}
\vspace{-0.2cm}
\includegraphics[scale=0.5]{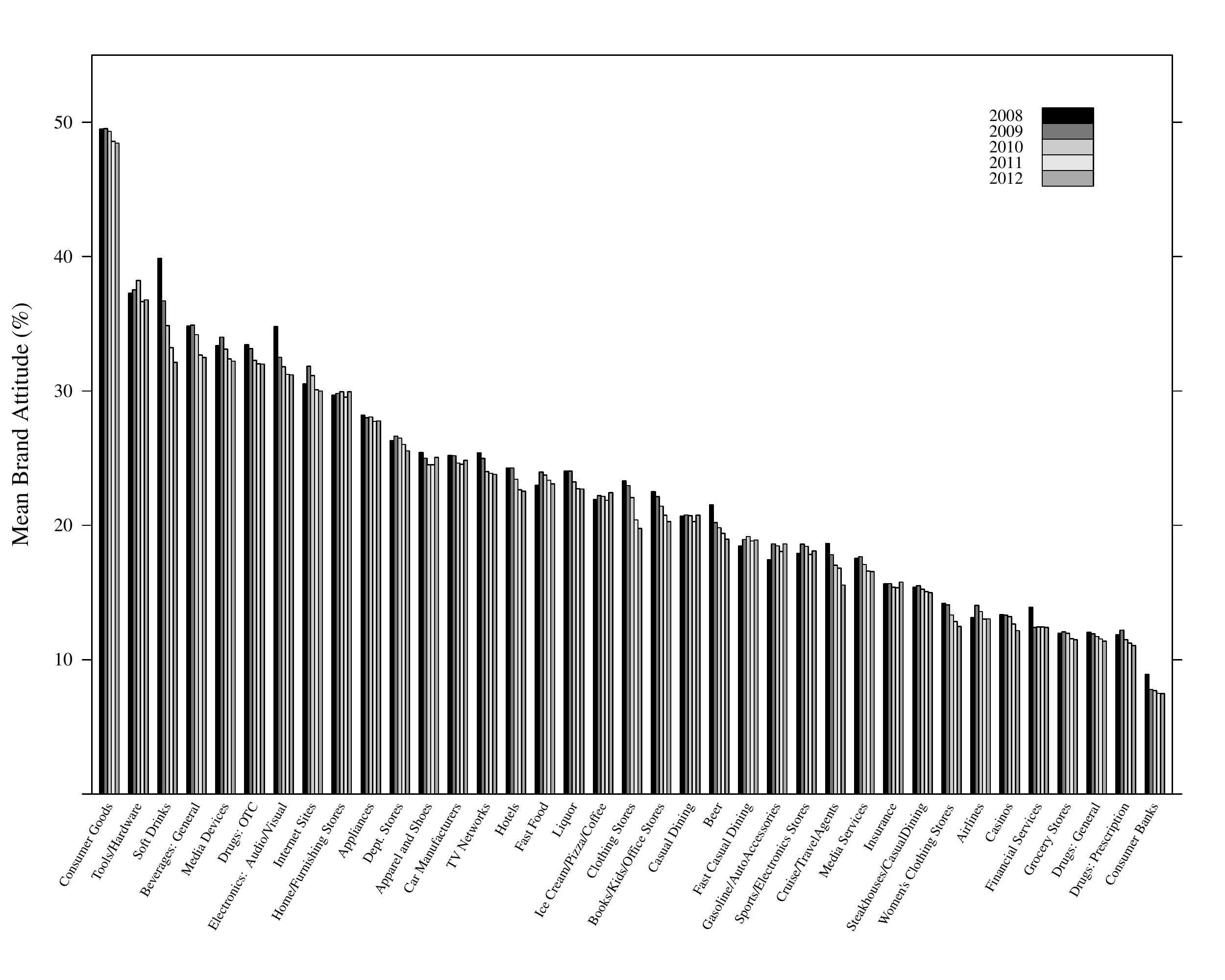}
\vspace{-0.4cm}
\end{center}
\end{figure}

Figure \ref{fig:MeanBA} shows how the average brand attitude metric (perceived quality, perceived value and recent satisfaction) changed for each industry in each year of the sample. The highest rated industries were consumer goods, tools/hardware and soft drinks; banking, prescription drugs, grocery retailing, casinos and financial service industries rank near the bottom. Some of these industry-level differences are partially driven by brand awareness, as large consumer goods brands are available throughout the U.S., whereas many brands in some of the lower-rated industries are more geographically dispersed (e.g., grocery retailers, consumer banks). Consumers in unserved regions would not indicate positive attitudes toward brands they have not encountered, as awareness must precede perceived quality, perceived value or recent satisfaction. 

The industry-level averages of brand attitude metrics are fairly stable across years in the sample, with a few exceptions. For example, consumer perceptions of soft drinks slipped sharply during the sample period. There is also a general negative trend in audio/visual electronics, though this shift in averages masks heterogeneity and consolidation; a few newer brands like LG and Acer improved, whereas some older brands (e.g., Kodak, Sony, Panasonic) fell. Despite these few exceptions, the industry-year averages were mostly stable during the sample period.

Table \ref{tb:brba} provides further information about how brand attitudes changed across years of the sample. Across all brands, perceived quality fell by an average of 1.6\% between 2008 and 2012; perceived value fell by 0.3\% and recent satisfaction fell by 0.9\%. These trends speak to the importance of controlling for time-varying unobservables in estimating relationships between ad spend and brand attitudes.

\renewcommand{\arraystretch}{1.1}
\begin{table}[htp]
\caption{Mean Brand Attitudes }
%\vspace{1em}
\label{tb:brba}
\footnotesize{
\begin{center}
{
\begin{tabular}{l|c|c|c|c|c|c}
\hline
& \multicolumn{2}{c|}{\textbf{Perceived Quality}}& \multicolumn{2}{c|}{\textbf{Perceived Value}}& \multicolumn{2}{c}{\textbf{Recent Satisfaction}}\\
\hline
& \textbf{In 2008} & \textbf{In 2012} & \textbf{In 2008} & \textbf{In 2012}  & \textbf{In 2008} & \textbf{In 2012} \\
\hline
\textbf{Mean} & 26.5\% & 24.9\% & 20.6\% & 20.3\% & 20.2\% & 19.3\% \\
\textbf{Median} & 24.4\% & 22.8\% & 17.4\% & 17.1\%  & 15.9\% & 15.0\% \\
%\textbf{Std.Dev.} & 16.5\% & 14.9\% & 4.45\% &13.9\% &13.1\% & 3.53\% & 15.5\% & 14.5\% & 3.54\%\\
\hline
\end{tabular}
}
\end{center}}
\end{table}

%\textcolor{red}{\st{National ad expenditures are positively correlated with local ad expenditures for 89\% of brands with median value of .24, but ad expenditures in other media do not present clear pattern. This may be because national and local ads focus on traditional media, and digital ads may be run separately from traditional ads.}}
%In sum, though correlations provide some directional evidence of positive relationship between brand attitudes and ad expenditures for some fraction of brands (60\%), but a substantial portion of brands still present negative correlation between brand attitudes and ad spending. Therefore, to evaluate whether brand attitudes can be advertising effectiveness metrics, a further modeling approach is required to control for alternative sources of time-varying and brand-specific fluctuations. \textcolor{red}{[DOES THIS CONTRADICT DESCRIPTIVE MODEL RESULTS IN 4.1?]}

\begin{figure}[htp]
\caption{Ad Spending by Industry and Year}
\label{fig:AdIndYr}
\begin{center}
\vspace{-0.2cm}
\includegraphics[scale=0.5]{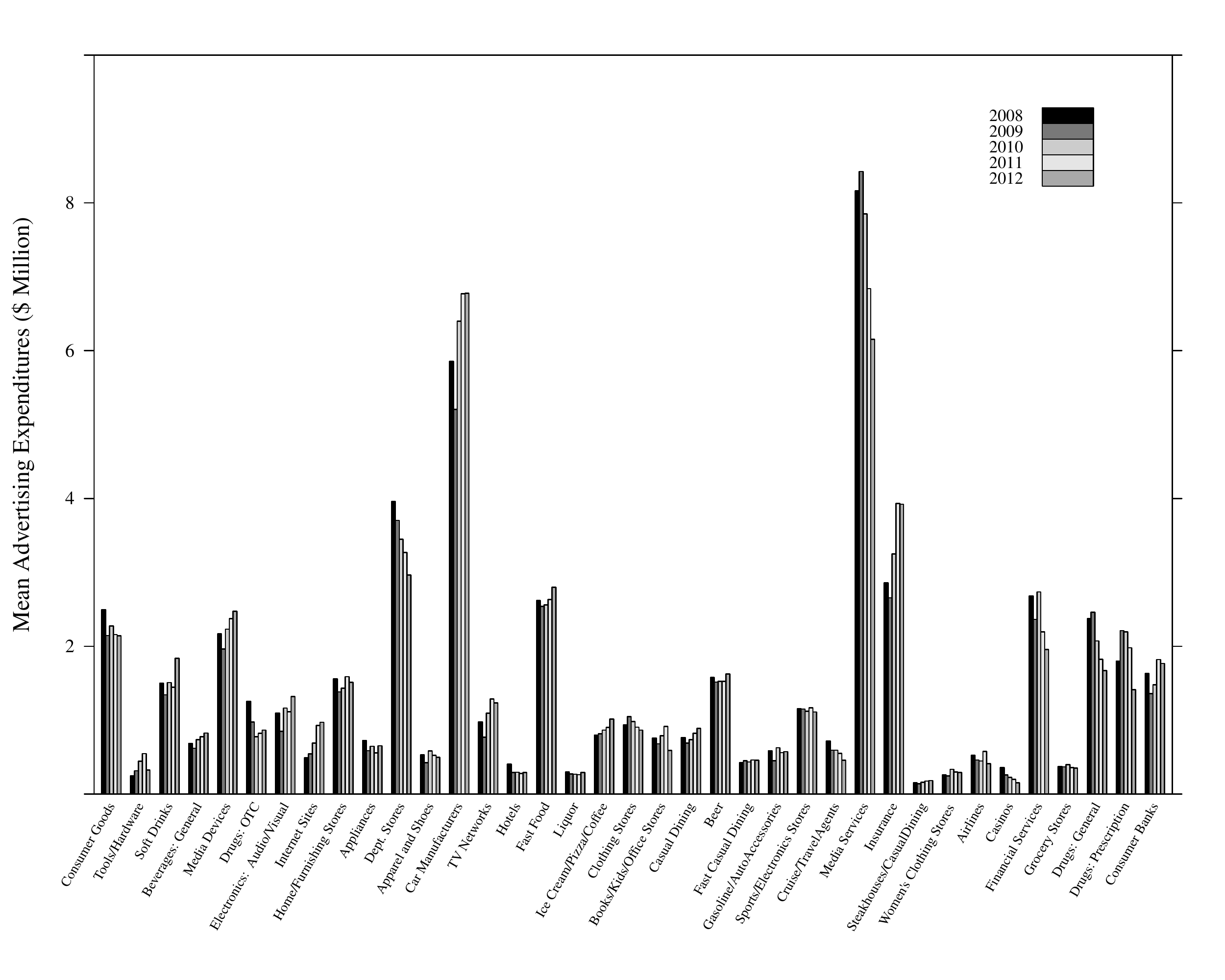}
\vspace{-0.4cm}
\end{center}
\end{figure}

Next, we summarize the ad spend data. Figure \ref{fig:AdIndYr} displays mean brand ad spend by industry and year. Media and automotive brands spent the most on advertising, followed by department store, insurance and quick service restaurant brands. Except for a few notable exceptions (e.g., media, department stores, insurance), most industries did not exhibit large changes in mean brand ad spend between 2008 and 2012. 

Table \ref{tb:brae} summarizes brand-level changes in ad spend by type of media between 2008 and 2012. There was some consolidation in national traditional ad spending during the sample, as the average brand's weekly expenditure rose by about \$50,000, whereas the median brand's weekly ad spend fell by \$22,000. Local traditional advertising fell substantially during the sample, with the average brand spending 25\% less per week in 2012 relative to 2008. Digital ads rose from \$110,000 per brand per week in 2008 to \$130,000 in 2012.

\renewcommand{\arraystretch}{1.1}
\begin{table}[htp]
\caption{Mean Advertising Expenditure (\$ Millions)}
%\vspace{1em}
\label{tb:brae}
\footnotesize{
\begin{center}
{
\begin{tabular}{l|c|c|c|c|c|c}
\hline
& \multicolumn{2}{c|}{\textbf{National Trad. Ads}}& \multicolumn{2}{c|}{\textbf{Local Trad. Ads}}& \multicolumn{2}{c}{\textbf{Digital Ads}}\\
\hline
& \textbf{In 2008} & \textbf{In 2012} & \textbf{In 2008} & \textbf{In 2012}  & \textbf{In 2008} & \textbf{In 2012} \\
\hline
\textbf{Mean} & \$1.02 & \$1.07 &  \$0.51 & \$0.38 & \$0.11 & \$0.13 \\
\textbf{Median} & \$0.27 & \$0.25 & \$0.11 & \$0.09 & \$0.01 & \$0.02 \\
%\textbf{Std.Dev.} & \$2.05 & \$2.17 &\$1.40 & \$0.84 & \$0.40 & \$0.35 \\
\hline
\end{tabular}
}
\end{center}}
\end{table}

\begin{landscape}
\begin{figure}[htp]
\caption{Weekly Brand Attitudes and Ad Expenditures of Selected Brands}
\label{fig:Evid}
\begin{center}
\vspace{-0.2cm}
{A. Toyota}\hspace{9.5cm}{B. Coke}
\includegraphics[scale=0.43]{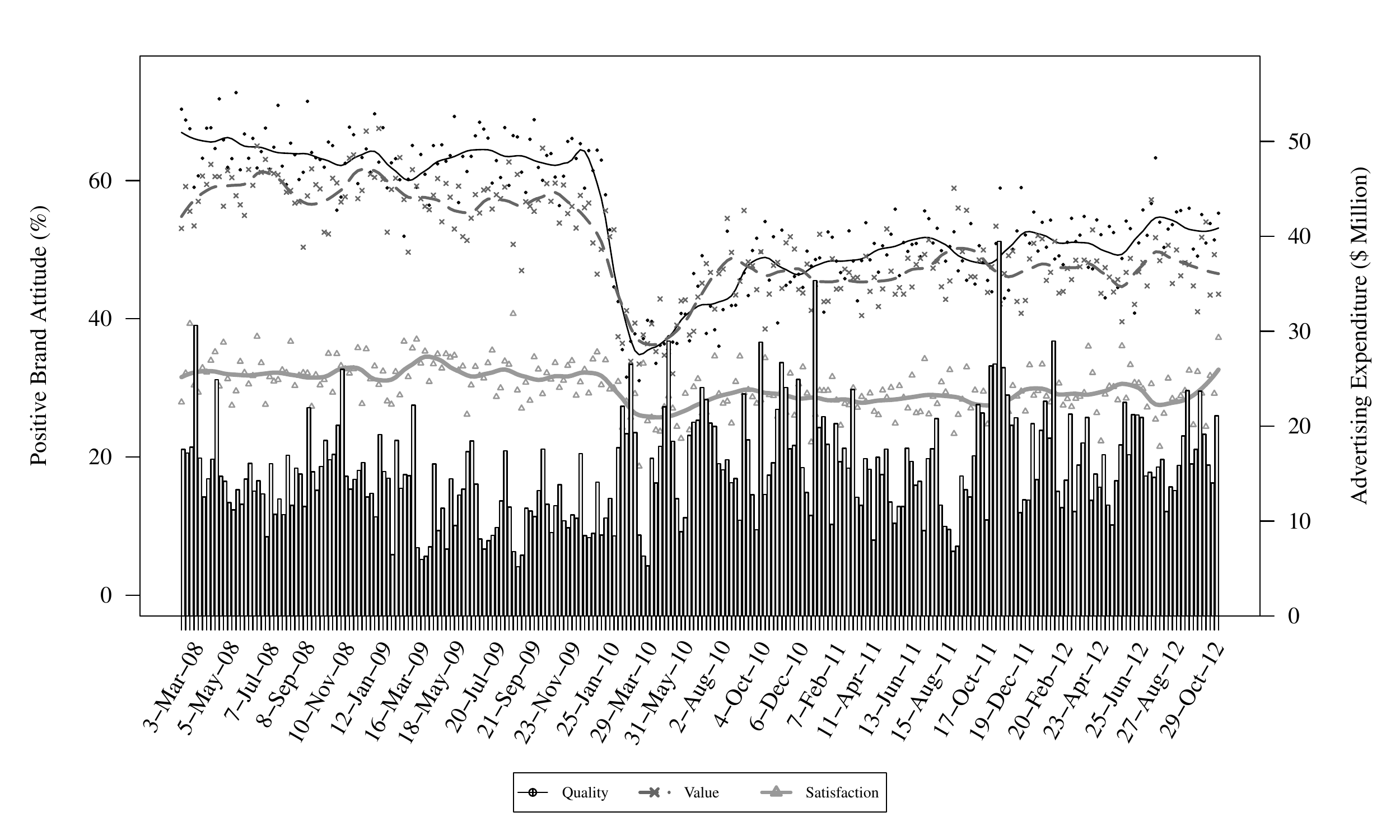}\includegraphics[scale=0.43]{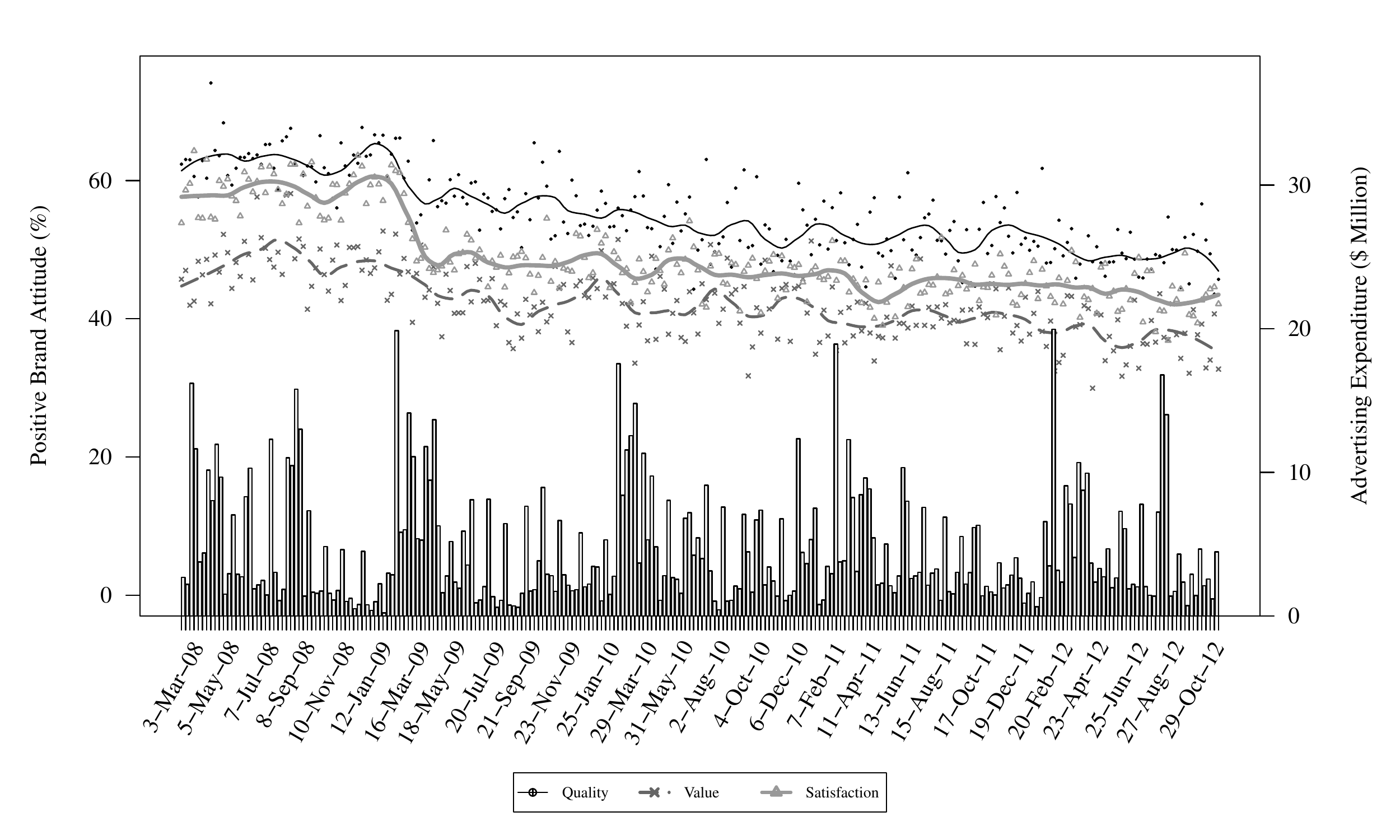}\\
{C. Ford}\hspace{9.5cm}{D. Apple}
\includegraphics[scale=0.43]{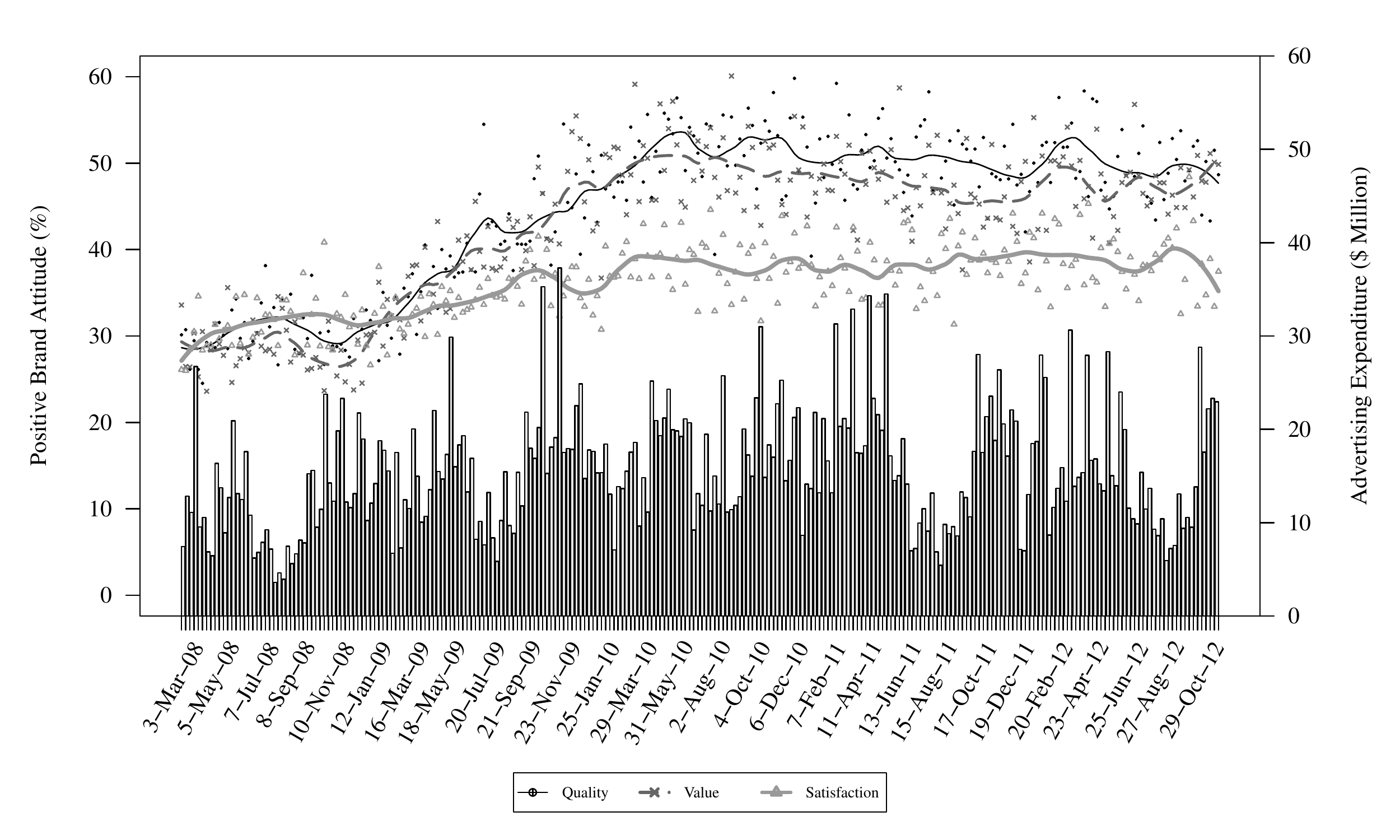}\includegraphics[scale=0.43]{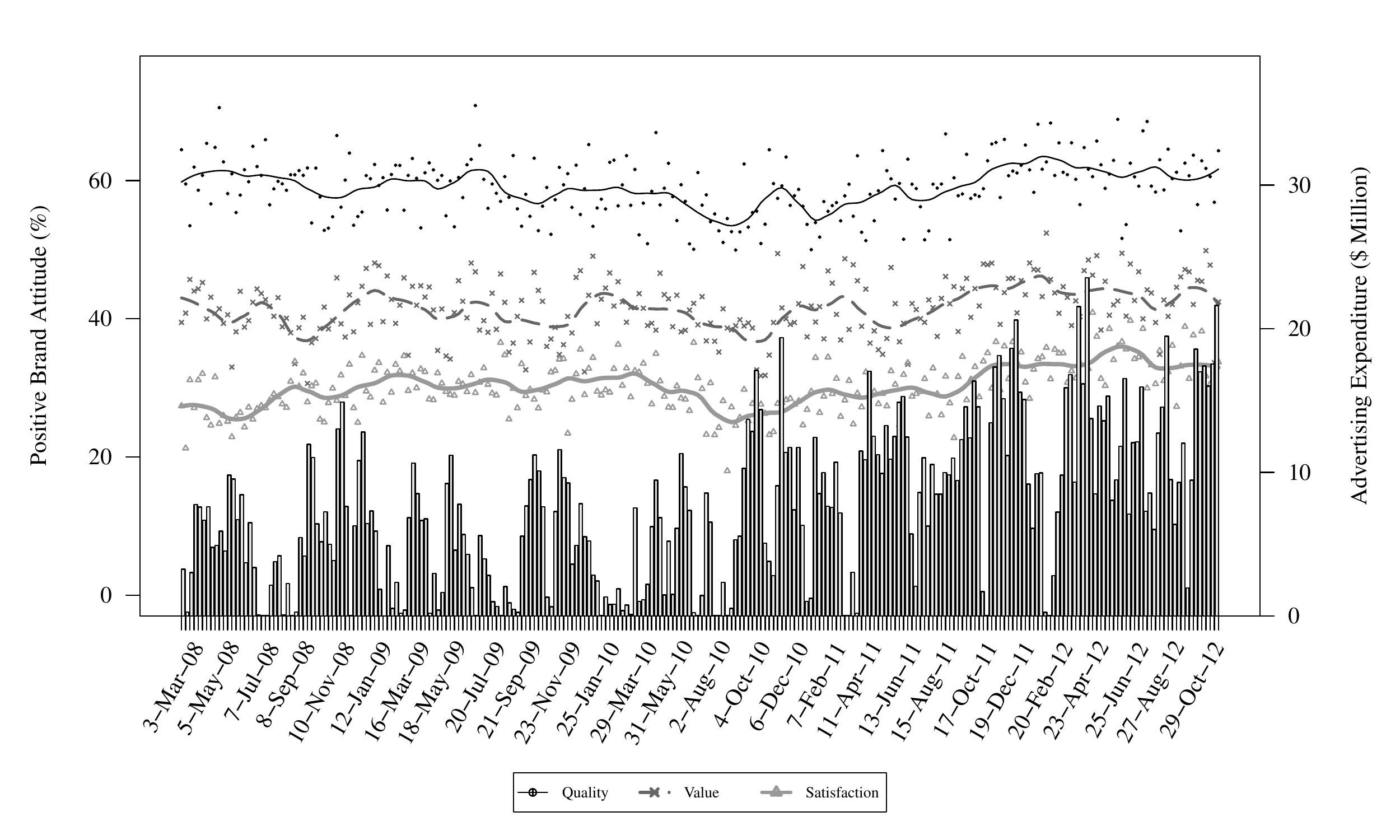}
\vspace{-0.4cm}
\end{center}
\end{figure}
\end{landscape}

Next, we focus on a series of visualizations for selected brands. Panel A of Figure \ref{fig:Evid} presents the three brand attitude metrics %\textcolor{red}{($qua_{bt}$, $val_{bt}$, and $sat_{bt}$)} 
for Toyota along with the brand's weekly ad spend data. % (i.e., sum of $n_{bt}$, $l_{bt}$, and $d_{bt}$). 
The dots represent brand attitude data points in each week whereas the trend lines are generated by local regression. Toyota's quality and value perceptions held steady at about 60\% until 2009, then dropped sharply to 30\% in 2010, due to a highly-publicized series of auto recall and safety incidents. They later began a slow recovery, though not quite up to the previous level. Unlike quality and value, recent satisfaction started out far lower around 30\%, and was less affected by the recall. Throughout this time period, advertising policy changed somewhat, but the peaks in ad spend do not correlate with immediate improvements in brand attitudes. 
  
Panel B of Figure \ref{fig:Evid} presents similar data for Coke. Like the aggregate trend of soft drinks in Figure \ref{fig:MeanBA}, and industry-level consumption figures more generally, respondents' attitudes toward Coke's perceived quality, perceived value, and recent satisfaction show gradual downward slopes which added up to a meaningful slide from 63\% to 50\% in perceived quality from 2008-2012. Although ad spend varies substantially throughout the sample, there is little visual evidence of any correspondence between brand attitudes and advertising expenditure.

Ford's brand attitude metrics, in Panel C, show a greater divergence than most brands. Quality and value perceptions increased early in the sample before leveling off around 50\%. Recent satisfaction initially approximated quality and value, but then leveled off at 40\%. Again, ad spend is highly variable, but there is little or no visual evidence that the peaks and troughs correspond to changes in brand attitudes. 

Finally, Apple's brand attitudes (Panel D) were fairly stable throughout the sample period. However, the three attitude metrics differed substantially, as perceived quality (about 60\%) was far higher than perceived value (about 40\%), which in turn substantially exceeded recent satisfaction (about 30\%). As in the other case studies, the brand's ad spend varied substantially, though it is again difficult to see a correspondence between advertising and brand attitudes. 

Several aspects of these four case studies replicate broadly throughout the sample. First, weekly brand attitude data generally vary around stable long-run averages. For this reason, we will use the number of survey respondents in each brand/week/metric observation to differentially weight observations according to how informative each observation is. Second, ad spend data are highly variable across weeks within each advertiser/quarter, though annual totals of advertising spend typically do not change much. Third, the relationship between advertising expenditure and brand attitudes is difficult to discern visually, even across long time horizons, motivating an econometric modeling approach to isolate the effects of advertising from possible confounding variables. Fourth, non-advertising events (such as the Toyota product harm crisis) may change brand attitudes significantly, suggesting a substantial importance of controlling for such confounds in a modeling framework.

Finally, we use the 252 weeks of data to construct brand-level correlations among the key variables. Table \ref{tb:brcor} presents the medians and central 90\% ranges of the brand-level correlations. Overall, brand attitude metrics are positively correlated with each other for most brands, with median correlations ranging from .10-.13. Among the ad spend variables, national and local traditional advertising are the most highly correlated (median of .24) whereas digital is weakly correlated with each (median .06 with national, .04 with local). Finally, confirming what we saw in the four brand case studies, the median correlations between brand attitudes and ad spend measures are all near zero, ranging from .00 (digital/satisfaction) to .03 (national/quality and national/value). 

\renewcommand{\arraystretch}{1}
\begin{table}[htp]
\caption{5\textsuperscript{th}, 50\textsuperscript{th}, and 95\textsuperscript{th} Percentiles among Brand-level Correlations}
%\vspace{1em}
\label{tb:brcor}
\begin{center}
{
\footnotesize{\tabcolsep=0.1cm\begin{tabular}{l|ccc|ccc}
\hline
& \textbf{Perceived Quality} & \textbf{Perceived Value} & \textbf{Recent Sat.} & \textbf{Nat'l Trad.} & \textbf{Local Trad.} & \textbf{Digital}\\
\hline
\textbf{Perceived Quality} & 1 & & & & & \\
\textbf{Perceived Value} & [-.04, .12, .49] & 1 & & & &  \\
\textbf{Recent Satisfaction} & [-.04, .13, .51] & [-.06, .10, .47] & 1  & & &\\
\hline
\textbf{National Trad.} & [-.14, .03, .22] & [-.12, .03, .22]& [-.13, .02, .20] & 1 & & \\
\textbf{Local Trad.} & [-.15, .01, .20] & [-.14, .02, .16] & [-.14, .01, .18] & [-.05, .24, .62] & 1 &  \\
\textbf{Digital } & [-.19, .01, .19] & [-.14, .01, .18] & [-.16, .00, .18]  & [-.09, .06, .38] & [-.10, .04, .33] & 1 \\
\hline
\end{tabular}}
}
\end{center}
\end{table}

\section{Endogeneity, Identifying Assumptions and Control Variables}
Numerous measurement and endogeneity problems arise in advertising response estimation. Traditional mass media advertisements are simultaneously transmitted to many people, either at the national or local level. Firms can often obtain noisy estimates of ad reach, and they can often directly measure or estimate the number of conversions (e.g. store visits, sales, leads accrued) that occurred after the message was transmitted. However, in the case of traditional advertising, they typically cannot link advertisement exposure with conversions at the individual level, as is often possible in digital advertising. In both traditional and digital advertising, it is difficult to separate advertising treatment effects from strategic targeting policies. In all cases, the fundamental difficulty is in determining what conversions would have occurred had the advertising not taken place. Estimation of weak advertising effects in statistically noisy environments is further complicated by frequent consumer disregard or avoidance of advertisements; repeated exposures and possibly nonlinear effects of ads on sales; frequent misattribution of advertised messages to competing brands, and other forms of competitive advertising interference; and advertisement copy rotation, ``wear-out'' and time-varying message effectiveness. %\footnote{Digital media, by contrast, enable observations of the intersection between exposure and conversion; but they typically do not lay bare the causal relationship without experimental randomization.}  

We are aware of three prominent research designs to estimate quasi-experimental advertising effects in traditional media. Each exploits particular institutional details:
\begin{itemize}
\item \cite{HK:2018} rely on local variation in regional preferences for watching featured sports teams, along with the simultaneous carriage of national ads in all local markets, and the allocation of ad slots to advertisers before the competing teams are known, to estimate the impact of national Super Bowl ads on local beverage sales. Under these conditions, each local market has a quasi-random component of its viewership of national ads, leading to exogenous variation in advertising exposures across local markets.
\item \cite{S:2018} exploits discontinuities in local television advertising intensity that occur at edges of contiguous geographic television markets to identify the treatment effects of local TV advertising on county-level response variables. The quasi-experimental logic relies on the similarity of neighboring counties on opposite sides of local television market borders, leading to numerous observations of county pairs exposed to different intensities of advertising treatment. The approach can be applied to estimate the effect of local TV advertising on any response indicator observed at the county level. 
\item \cite{LTW:2015} rely on quasi-random national TV advertisement insertion times to estimate effects on brand website traffic and sales. The treatment/control logic relies on examining narrow windows of time, such as two minutes immediately before the TV ad and two minutes after, along with typical TV industry practices of contractually unspecified commercial break start times and randomized advertisement ordering within commercial breaks. The treatment/control logic assumes that the observation windows are narrow enough that no competing explanations can plausibly account for changes in pre-ad and post-ad response variables.
\end{itemize}
Each of these research designs advances our ability to estimate causal advertising effects by applying quasi-experimental econometric techniques to retrospective field data, but each relies on specific institutional details. In particular, none of these strategies is able to answer the research questions that motivate the current analysis, as brand attitude data are only observed for the national market on a weekly basis for each brand.

%These approaches are promising in the sense that they seek to exploit quasi-random variation to estimate advertising effects in quasi-experimental research designs. It would be possible to combine some strategies, although we are not aware of any existing attempts to do so. It is also possible to use instrumental variables in some cases, although such exercises do not seem to be numerous, perhaps because good instrumental variables are hard to find. In conclusion, there are many marketers that might not be able to apply any of these strategies, or for whom these strategies might not yield precise estimates. Yet these marketers still would like to use field data to understand effects of advertising.

In contrast to academic research, practitioners often identify advertising effects using an assumption of {\it precedence}.\footnote{See, for example, http://pages.stern.nyu.edu/\~atakos/studentevents/3-28-12MeasuringROMISlideDeck.pdf or https://en.wikipedia.org/wiki/Marketing\_mix\_modeling, accessed March 2018.}
That is to say, if advertising preceded sales, then any discernible response of sales is attributed to the advertising that came before it. There are also some published studies of advertising effects that infer causality using a similar identifying assumption. % (e.g. Naik et al. \citeyear{NPS:2008}, Bruce et al. \citeyear{BPN:2012}). 
The typical argument for the validity of this identification strategy is that a brand's ad spending must be determined prior to the firm's observation of the response variable.

One need not presume much sophistication on the part of a marketer to show that the precedence assumption can be tainted by unobserved variables. As a simple example, suppose that a brand knows that demand tends to rise in a promotion week, and that the brand prefers to advertise more heavily during periods of peak demand; then both sales and advertising could be simultaneously influenced by the third variable (promotion week), yielding a spurious or inflated finding of ad effects on sales. Similar arguments can be based on any number of unobserved variables---e.g., changes in wholesale or retail prices, distribution, product assortments and line extensions, trade promotions, competitor marketing mix variables---that may correlate with both advertising and sales. 

Arguments against precedence need not depend on unobserved variables. For example, if the marketer correctly anticipated a likely future change in future revenues, and set ad spend as a proportion of anticipated future revenues, then ad effect estimates may be biased upward by simultaneity. The key problem is that the advertising policy function is unobserved by the econometrician and may depend on anticipation of future changes in the response variable.%\footnote{In some cases, such concerns might be partially mitigated by inclusion of lagged values of the response variable in the model, as they may predict future . However, we view that approach as a partial remedy at best.}

Anecdotally, when we discuss such issues with practitioners, we find three typical reactions. One is an understanding and agreement that advertising response estimates are likely to be biased, coupled with a belief that biased estimates are likely better than no estimates at all. Another common response is a gap in understanding endogeneity issues: we rationalize this with the observation that most business schools did not start teaching causal methods until relatively recently; large brands have traditionally not screened their marketing recruits for this skill; and incentives to experiment may be distorted by the principal/agent relationships that are nearly ubiquitous in practice. The third common refrain is a deep skepticism that brand advertising decisions are made strategically. Executives in several organizations have told us that their company sets quarterly or annual advertising budgets and that the agencies allocate the budget across media programs and weeks without anticipation of likely changes in the market.\footnote{We remain circumspect about this argument, as agencies may be aware of their clients' evaluation function and act to maximize their own incentives to demonstrate advertising effects to their clients.}

Naturally, we are unable to characterize the full set of endogeneity problems for the 575 brands and 37 different industries represented in these data. Yet we would like to consider how various control strategies might influence estimates of advertising effects on brand attitudes in a broad sample of mature brands that advertise regularly. We consider four main specifications:
\begin{enumerate}
\item{Descriptive regression with multiple lags of brand advertising and competitor advertising, controlling for lagged response variables, brand effects, time effects and weighted standard errors.}
\item{Descriptive regression (1) with industry/year/week fixed effects added (we call these ``industry/week'' effects).}
\item{Descriptive regression (1) with brand/year/quarter fixed effects added (we call these ``brand/quarter'' effects).}
\item{Descriptive regression (1) with both industry/week and brand/quarter fixed effects added, 21,194 fixed effects in all.}
\end{enumerate}

The industry/week fixed effects should control for any industry-level unobservables in a given week that affect all brands' advertising expenditures, such as seasonal fluctuations in industry demand. There are many brands observed within every industry, providing sufficient variation to estimate a separate industry fixed effect for each week of the sample. 

The brand/quarter fixed effects should control for any brand-level unobservables that persist across weeks within a quarter, such as budgetary changes or persistent changes in unobserved marketing variables. There are 13 weeks of brand attitude data within each quarter, yielding sufficient data to estimate a separate brand fixed effect for each quarter in the sample. 

In the model that contains both industry/week and brand/quarter fixed effects, the assumption required for causal interpretation is that brand-week fluctuations of ad spending are uncorrelated with ({\it i}) brand-week departures from brand-quarter unobservables and ({\it ii}) brand-week departures from industry-week unobservables. Although still imperfect, this assumption is 	much weaker and more plausible than the typical assumption that advertising spend is uncorrelated with brand-week unobservables. We think this assumption is probably reasonable for most brands whose attitudes are largely stable across quarters, as is typical in the sample that we study. However, we acknowledge that the assumption may be violated, especially in the presence of systematic weekly fluctuations in drivers of brand attitudes that can be anticipated by the brand and are used to set advertising policies. 

%\textcolor{red}{Max: In revision strategy, we have "Estimate a log-log model including fixed effects for category-week interactions and brand-quarter interactions (about 21,194 fixed effects in total), along with 13 lags of all brand metrics and adspend, and use brand-week survey sample size to weight the standard errors"}

Of course, what we would really like to control for is brand/week fixed effects, but these would covary perfectly with the advertising data and therefore would prevent estimation of the quantities of primary interest. Still, we believe that the two sets of control variables might, together, handle some common sources of endogeneity and let us offer, at minimum, a first approximation of the effects of ad spend on brand attitudes. We also think it might be instructive to observe how the control variables change the qualitative conclusions.

In sum, we try to control for time-varying confounds as much as possible, so the main model results can be interpreted as causal subject to a clearly specified identifying assumption. %To obtain fully causal estimates, one should ideally run an experiment; however, advertisers seem to be surprisingly experiment-averse,\footnote{\textcolor{red}{1 sentence summarizing \cite{RS:2016}}} and large sample sizes may be needed to adequately power advertising experiments.\footnote{\textcolor{red}{1 sentence summarizing \cite{LR:2014}}} 
As we await highly powered RCTs in traditional media or more comprehensive quasi-experimental research designs, we hope that the estimates below may be viewed as suggestive of causal effects, subject to appropriate caveats, and possibly useful to marketers and their advertising agencies as they think about how to allocate advertising budgets and apply appropriate control variables in similar settings.

\section{Models}

The main goals of this paper are to estimate relationships between brand attitudes and ad spend variables; to show how these effects vary across types of advertising media; and to illustrate how control variables change the estimates. We seek to ``let the data speak'' by specifying simple models and contrasting the results across comparable metrics and control variables.

We represent the log of one plus any focal brand attitude metric for brand $b$ in industry $i$ in week $t$ as $y_{bt}$ and the other two metrics with $y^{\prime}_{bt}$ and $y^{\prime \prime}_{bt}$. The log of one plus national traditional, local traditional and digital ad spend for brand $b$ in week $t$ are $na_{bt}$, $la_{bt}$ and $da_{bt}$, respectively; its competitors' log of one plus ad spend observed in week $t$ are $na^c_{bt}$, $la^c_{bt}$, and $da^c_{bt}$ in national, local and digital media, respectively.%\footnote{Following standard practice, we add one to all observed values of ad spend to avoid taking logs of zero.}

The model specification is 
\begin{align} \label{eq:logqua}
y_{bt} =& \sum_{\tau=1}^{T_y}{\alpha^{y}_{\tau}{y}_{b,t-\tau}}+\sum_{\tau=1}^{T_y}{\alpha^{y\prime}_{\tau}{y^{\prime}}_{b,t-\tau}}+\sum_{\tau=1}^{T_y}{\alpha^{y^{\prime \prime}}_{\tau}{y^{\prime \prime}}_{b,t-\tau}} \nonumber\\ 
& + \sum_{\tau=0}^{T_a}{\beta^{ny}_{\tau}{na}_{b,t-\tau}}+\sum_{\tau=0}^{T_a}{\beta^{ly}_{\tau}{la}_{b,t-\tau}}+\sum_{\tau=0}^{T_a}{\beta^{dy}_{\tau}{da}_{b,t-\tau}} \\
& + \sum_{\tau=0}^{T_a}{\beta^{ny,c}_{\tau}{na}^c_{b,t-\tau}}+\sum_{\tau=0}^{T_a}{\beta^{ly,c}_{\tau}{la}^c_{b,t-\tau}}+\sum_{\tau=0}^{T_a}{\beta^{dy,c}_{\tau}{da}^c_{b,t-\tau}}\nonumber\\
& +Z_{bt}\Theta^y +\epsilon^{y}_{bt}\nonumber.
\end{align}

\noindent The number of lags of attitude metrics is held constant at $T_y=13$. The model also includes $T_a=5$ lags of each advertising variable, on the theory that the direct effects of advertising on brand attitudes seem unlikely to persist beyond five weeks. The qualitative results change remarkably little with $T_a$, as shown in Table \ref{tb:diff_lagsqua} in the appendix. 

$Z_{bt}$ specifies the vector of fixed effects. The baseline specification includes fixed effects for each brand in the sample and for each week in the sample. Subsequent regressions also include industry/week interactions; brand/quarter interactions; and both industry/week and brand/quarter interactions.

We use the number of survey respondents for brand attitude $y$ in week $t$ to weight the standard errors, as brand attitudes based on larger sample sizes are more informative. Parameters for each brand attitude model $y$ are estimated by minimizing $\{\left(\mathbf{N^y}\mathbf{E^y}\right)'\cdot\left(\mathbf{N^y}\mathbf{E^y}\right)\}$, where $\mathbf{N^y}=[n^y_{bt}]$, $n^y_{bt}$ is the number of survey respondents for brand attitude question $y$ for brand $b$ in week $t$, and $\mathbf{E^y}=[\epsilon^y_{bt}]$. %By placing greater weight on the brand/week observations with larger numbres of survey respondents, parameter estimates can be made more precise. 

\section{Findings}\label{s:findings}

We start by comparing fit statistics across models (Table 5). The descriptive model explains the large majority of variation in the brand attitude data, with adjusted R-squared statistics ranging from .955 to .978. These high model fit statistics are to be expected, as the brand attitude data are strongly autocorrelated, and the baseline specification includes lagged brand attitudes in addition to the 575 brand fixed effects and 252 week fixed effects. 

\begin{table}%[tb]
\caption{Model Comparison with Different Control Variables}
%\vspace{1em}
\label{tb:mcomp}
\begin{center}{
\scriptsize{
\begin{tabular}{ll|cccc}																
\hline											
 & {Model}			&	{Descriptive}	&	{Ind./Wk.}	&	{Br./Qtr.}	&	{All Controls}\\
\hline											
Adjusted	&	Perceived Quality	&	.960	&	.968	&	.964	&	.971	\\
R-Squared &	Perceived Value	&	.955	&	.962	&	.959	&	.966	\\
	&	Recent Satisfaction	&	.978	&	.982	&	.980	&	.983	\\
\hline											
Model&	Lagged Attitudes, &	Yes	&	Yes	&	Yes	&	Yes	\\
Includes... &	Advertising Variables, & & & & \\	%Yes	&	Yes	&	Yes	&	Yes	\\
	&	Brand and Week Effects & & & & \\	%Yes	&	Yes	&	Yes	&	Yes	\\
%	&	Week	&	Yes	&	Yes	&	Yes	&	Yes	\\
	&	Industry/Week Effects	&	No	&	Yes	&	No	&	Yes	\\
	&	Brand/Quarter	Effects &	No	&	No	&	Yes	&	Yes	\\
\hline																																											
\end{tabular}}}																		
\end{center}
\end{table}	

The second column of Table 5 displays the adjusted R-squared statistics when the 9,324 industry/week fixed effects are added to the descriptive model. Even though the fit statistics penalize the large increase in model complexity, the proportion of unexplained variance falls substantially, from .022-.045 in the descriptive model, to .018-.038 in the model with industry/week controls. The F-statistic rejects the null hypothesis that industry/week fixed effects should be excluded from the model ($p<.001$).

Similarly, the brand/quarter fixed effects reduce the proportion of unexplained variance from .022-.045 in the descriptive model to .020-.041, even after penalizing for the additional 11,500 parameters. The F-statistic rejects the null hypothesis that brand/quarter fixed effects should be excluded from the model ($p<.001$).

Finally, the model that includes both brand/quarter and industry/week fixed effects further reduces the proportion of unexplained variance, relative to each of the models with only a single set of control variables. The F-statistics reject the null hypotheses that industry/week fixed effects or brand/quarter fixed effects should be excluded from the model ($p<.001$), regardless of whether the other set of control variables is included in the baseline model or not.

The data show that, despite the limited room to improve on the descriptive model, each set of control variables is {\it individually and jointly} important for explaining brand attitudes. Of course, model fit statistics do not prove that the parameter estimates are unbiased or even that the results make sense. Next, we interpret and contrast the findings of the descriptive model and the all-controls model. Results from models that include industry/week controls only, and brand/quarter controls only, are provided in the appendix.

%\subsection{Qualitative Findings}\label{ss:findings}

Tables \ref{tb:adest_qua}$\sim$\ref{tb:adest_sat} provide all advertising parameter estimates from the descriptive and the all-controls specifications, for each of the brand attitude models and for each type of advertising, contrasting each brand's own advertising effects with its competitors' advertising effects. The estimates of lagged brand attitudes within each model are presented in the appendix. Overall, although some of the descriptive model results are intuitive, many of them are quite challenging to interpret. In contrast, the all-controls model advertising parameter estimates are more logical and more coherent ad effects, providing some reassurance that they may be closer to the true causal effects. Parameter estimate precision is indicated by stars for significance levels; standard errors are provided in the appendix.

\renewcommand{\arraystretch}{1}
\begin{table}%[tb]
\caption{Ad Parameter Estimates for Perceived Quality}
%\vspace{1em}
\label{tb:adest_qua}
\begin{center}
{
\scriptsize{
\tabcolsep=0.15cm
\begin{tabular}{ll|rlrl|rlrl}
\hline																			
\multicolumn{2}{l|}{Specifications}			&	\multicolumn{4}{c|}{Descriptive Model}							&	\multicolumn{4}{c}{All Controls Model}							\\
\hline																			
\multicolumn{2}{l|}{Ad Expenditures}			&	\multicolumn{2}{c}{Own}			&	\multicolumn{2}{c|}{Comp.}			&	\multicolumn{2}{c}{Own}			&	\multicolumn{2}{c}{Comp.}			\\
\hline																			
National	&	($\tau=0$)	&	4.12E-05	&		&	8.67E-05	&		&	2.27E-05	&		&	-3.93E-04	&		\\
Trad. Ads	&	($\tau=1$)	&	1.04E-04	&	**	&	2.96E-04	&	**	&	6.67E-05	&	**	&	-4.04E-04	&		\\
	&	($\tau=2$)	&	-2.12E-05	&		&	-7.80E-05	&		&	2.66E-05	&		&	-3.43E-04	&		\\
	&	($\tau=3$)	&	2.63E-05	&		&	-1.78E-04	&		&	3.26E-05	&		&	-4.89E-04	&	*	\\
	&	($\tau=4$)	&	4.13E-05	&		&	1.67E-04	&		&	6.99E-05	&	**	&	-9.75E-05	&		\\
	&	($\tau=5$)	&	-2.60E-05	&		&	-5.27E-05	&		&	4.81E-05	&	*	&	2.10E-05	&		\\
\hline																			
Local	&	($\tau=0$)	&	3.91E-05	&		&	8.32E-05	&		&	2.50E-05	&		&	1.35E-04	&		\\
Trad. Ads	&	($\tau=1$)	&	5.70E-05	&	*	&	-4.60E-05	&		&	6.54E-05	&	**	&	2.21E-04	&		\\
	&	($\tau=2$)	&	2.68E-05	&		&	2.40E-04	&	**	&	-8.57E-06	&		&	-5.40E-04	&	*	\\
	&	($\tau=3$)	&	1.29E-05	&		&	7.30E-05	&		&	3.51E-05	&		&	1.81E-04	&		\\
	&	($\tau=4$)	&	1.70E-05	&		&	5.90E-05	&		&	2.68E-05	&		&	-1.16E-04	&		\\
	&	($\tau=5$)	&	-2.68E-06	&		&	-1.37E-04	&		&	3.97E-05	&		&	-3.72E-04	&		\\
\hline																			
Digital	&	($\tau=0$)	&	4.29E-05	&		&	4.72E-05	&		&	2.89E-05	&		&	1.97E-04	&		\\
Ads	&	($\tau=1$)	&	2.20E-05	&		&	2.08E-04	&		&	6.17E-05	&		&	5.25E-04	&		\\
	&	($\tau=2$)	&	-3.48E-05	&		&	-3.26E-04	&	*	&	-3.69E-05	&		&	-2.66E-04	&		\\
	&	($\tau=3$)	&	3.61E-05	&		&	1.23E-04	&		&	3.40E-05	&		&	8.08E-05	&		\\
	&	($\tau=4$)	&	1.54E-06	&		&	3.19E-04	&	*	&	-4.94E-06	&		&	-1.37E-04	&		\\
	&	($\tau=5$)	&	-5.70E-05	&		&	-3.83E-04	&	**	&	6.46E-06	&		&	-7.89E-04	&	*	\\
\hline																			
\multicolumn{10}{l}{* Significant at the 95\% confidence level.}\\												
\multicolumn{10}{l}{** Significant at the 99\% confidence level.}												\end{tabular}}}
\end{center}
\end{table}

The descriptive model results in the first column of Table \ref{tb:adest_qua} show that perceived quality tends to increase with one-week lags of own ad spend in both national and local traditional media, which is not very surprising. However, the second column indicates that perceived quality also increases with competitors' ad spend in both national and local traditional media; and it both increases and decreases with various lags of competitors' digital ad spend. It certainly could be possible for a brand's attitude metrics to increase with competitors' advertising, for example, if competitors' ads draw new consumers to the category who engage in search to discover multiple brands' offerings and attributes. However, within the context of mature brands that advertise regularly, this would seem like an incongruous finding, as most of the brands in this study are already widely known at the start of the sample period. Our expectation prior to conducting this research was that brand attitudes were likely to fall, or at least not increase, with competitors' ad spend.

The all-controls model advertising parameter estimates in the third column of Table \ref{tb:adest_qua} also show that perceived quality tends to increase with ad spend in national and local traditional media, but parameter estimates are more precise. The fourth column shows that perceived quality tends to decrease with competitors' ad spend all three types of media, which is more consistent with our expectation in the category of mature brands with regular advertising.														

Table \ref{tb:adest_val} presents a similar contrast in perceived value. The descriptive model shows that perceived value both increases {\it and} decreases with lagged advertising in national traditional media. In fact, this confusing pattern of both positive/significant {\it and} negative/significant results appears in {\it all} competitor-advertising/media combinations with multiple significant findings in the descriptive model. Such results are quite difficult to interpret and seem to cast doubt on the validity of the findings. However, the all-controls model shows no such dissonance; perceived value increases with advertisers' own ad spend in all three types of media, decreases with competitors' ad spend in local traditional media, and (surprisingly) increases with competitors' lagged digital spend.

\renewcommand{\arraystretch}{1}
\begin{table}%[tb]
\caption{Ad Parameter Estimates for Perceived Value}
%\vspace{1em}
\label{tb:adest_val}
\begin{center}
{
\scriptsize{
\tabcolsep=0.15cm
\begin{tabular}{ll|rlrl|rlrl}
\hline																			
\multicolumn{2}{l|}{Specifications}			&	\multicolumn{4}{c|}{Descriptive Model}							&	\multicolumn{4}{c}{All Controls Model}							\\
\hline																			
\multicolumn{2}{l|}{Ad Expenditures}			&	\multicolumn{2}{c}{Own}			&	\multicolumn{2}{c|}{Comp.}			&	\multicolumn{2}{c}{Own}			&	\multicolumn{2}{c}{Comp.}			\\
\hline																			
National	&	($\tau=0$)	&	4.00E-05	&	*	&	2.78E-04	&	**	&	3.26E-05	&		&	1.32E-04	&		\\
Trad. Ads	&	($\tau=1$)	&	6.71E-05	&	**	&	1.10E-04	&		&	7.43E-05	&	**	&	-3.70E-04	&		\\
	&	($\tau=2$)	&	5.80E-05	&	**	&	-1.51E-04	&		&	6.41E-05	&	**	&	-1.84E-04	&		\\
	&	($\tau=3$)	&	4.96E-06	&		&	-9.66E-05	&		&	1.88E-05	&		&	-3.06E-04	&		\\
	&	($\tau=4$)	&	-1.71E-05	&		&	-1.93E-04	&		&	3.22E-05	&		&	-6.98E-05	&		\\
	&	($\tau=5$)	&	-5.56E-05	&	**	&	-2.19E-04	&	*	&	-1.15E-05	&		&	-3.53E-04	&		\\
\hline																			
Local	&	($\tau=0$)	&	3.14E-05	&		&	-1.49E-04	&		&	2.91E-05	&		&	-3.25E-04	&		\\
Trad. Ads	&	($\tau=1$)	&	2.91E-05	&		&	-1.27E-04	&		&	3.60E-05	&		&	-4.28E-04	&	*	\\
	&	($\tau=2$)	&	8.30E-06	&		&	2.41E-04	&	**	&	2.66E-05	&		&	4.22E-05	&		\\
	&	($\tau=3$)	&	-1.88E-08	&		&	-2.05E-04	&	*	&	1.03E-05	&		&	-2.41E-04	&		\\
	&	($\tau=4$)	&	2.60E-07	&		&	1.29E-04	&		&	7.87E-06	&		&	5.94E-05	&		\\
	&	($\tau=5$)	&	4.31E-05	&		&	-4.77E-05	&		&	7.78E-05	&	**	&	1.50E-04	&		\\
\hline																			
Digital	&	($\tau=0$)	&	8.18E-05	&	*	&	3.33E-05	&		&	2.50E-05	&		&	-5.36E-04	&		\\
Ads	&	($\tau=1$)	&	-4.19E-05	&		&	2.67E-04	&	*	&	-5.35E-05	&		&	-3.51E-04	&		\\
	&	($\tau=2$)	&	7.19E-05	&		&	-2.63E-04	&	*	&	8.18E-05	&	*	&	3.85E-04	&		\\
	&	($\tau=3$)	&	-6.42E-05	&		&	7.25E-05	&		&	-5.54E-05	&		&	2.62E-05	&		\\
	&	($\tau=4$)	&	3.68E-05	&		&	-2.03E-04	&		&	2.82E-05	&		&	-1.50E-04	&		\\
	&	($\tau=5$)	&	4.70E-06	&		&	3.83E-04	&	**	&	4.50E-05	&		&	7.99E-04	&	**	\\
\hline																			
\multicolumn{10}{l}{* Significant at the 95\% confidence level.}\\																			
\multicolumn{10}{l}{** Significant at the 99\% confidence level.}												
\end{tabular}}}
\end{center}
\end{table}

\renewcommand{\arraystretch}{1}
\begin{table}%[tb]
\caption{Ad Parameter Estimates for Recent Satisfaction}
%\vspace{1em}
\label{tb:adest_sat}
\begin{center}
{
\scriptsize{
\tabcolsep=0.15cm
\begin{tabular}{ll|rlrl|rlrl}
\hline																			
\multicolumn{2}{l|}{Specifications}			&	\multicolumn{4}{c|}{Descriptive Model}							&	\multicolumn{4}{c}{All Controls Model}							\\
\hline																			
\multicolumn{2}{l|}{Ad Expenditures}			&	\multicolumn{2}{c}{Own}			&	\multicolumn{2}{c|}{Comp.}			&	\multicolumn{2}{c}{Own}			&	\multicolumn{2}{c}{Comp.}			\\
\hline																			
National	&	($\tau=0$)	&	2.72E-05	&		&	-9.31E-05	&		&	2.05E-05	&		&	-9.31E-05	&		\\
Trad. Ads	&	($\tau=1$)	&	2.37E-06	&		&	-7.33E-05	&		&	1.72E-05	&		&	7.75E-05	&		\\
	&	($\tau=2$)	&	2.58E-05	&		&	-2.96E-06	&		&	3.98E-05	&	*	&	-1.60E-05	&		\\
	&	($\tau=3$)	&	2.03E-05	&		&	-2.96E-04	&	**	&	4.24E-05	&	**	&	-7.59E-06	&		\\
	&	($\tau=4$)	&	1.79E-05	&		&	1.54E-04	&	*	&	3.94E-05	&	*	&	1.01E-04	&		\\
	&	($\tau=5$)	&	-2.32E-05	&		&	-4.45E-05	&		&	9.08E-06	&		&	-5.57E-05	&		\\
\hline																			
Local	&	($\tau=0$)	&	-9.36E-07	&		&	4.35E-05	&		&	-2.00E-05	&		&	-3.31E-04	&	*	\\
Trad. Ads	&	($\tau=1$)	&	3.04E-05	&		&	7.80E-05	&		&	1.34E-05	&		&	-2.75E-04	&		\\
	&	($\tau=2$)	&	-2.54E-05	&		&	-1.47E-05	&		&	-2.06E-05	&		&	-7.14E-05	&		\\
	&	($\tau=3$)	&	-1.39E-05	&		&	1.92E-04	&	**	&	-7.85E-06	&		&	2.14E-04	&		\\
	&	($\tau=4$)	&	-8.51E-06	&		&	3.08E-05	&		&	-5.97E-06	&		&	-1.12E-04	&		\\
	&	($\tau=5$)	&	4.83E-06	&		&	-5.61E-05	&		&	1.49E-05	&		&	-1.15E-04	&		\\
\hline																			
Digital	&	($\tau=0$)	&	1.26E-05	&		&	-4.48E-05	&		&	2.94E-06	&		&	7.93E-05	&		\\
Ads	&	($\tau=1$)	&	2.21E-05	&		&	-7.25E-05	&		&	3.04E-05	&		&	-2.02E-04	&		\\
	&	($\tau=2$)	&	1.31E-06	&		&	1.56E-04	&		&	-6.54E-07	&		&	4.28E-04	&		\\
	&	($\tau=3$)	&	4.51E-06	&		&	1.86E-04	&		&	-2.89E-07	&		&	-3.62E-04	&		\\
	&	($\tau=4$)	&	9.13E-06	&		&	-1.16E-04	&		&	-8.46E-06	&		&	-2.58E-04	&		\\
	&	($\tau=5$)	&	4.11E-06	&		&	-1.62E-04	&		&	-3.37E-07	&		&	-7.72E-04	&	**	\\
\hline																			
\multicolumn{10}{l}{* Significant at the 95\% confidence level.}\\																			
\multicolumn{10}{l}{** Significant at the 99\% confidence level.}												\end{tabular}}}
\end{center}
\end{table}

All-controls model estimates in Table \ref{tb:adest_sat} show that recent satisfaction rises with the own advertising in national traditional media, but does not seemingly react to local or digital ad spend. However, competitors' ad spend on local traditional and digital ads seem to harm recent satisfaction attitudes.

%The top panel of Table \ref{tb:adest} shows that, although some of the descriptive model results are intuitive, many of them are quite challenging to interpret. For example, it is not very surprising that the first column shows that perceived quality tends to increase with one-week lags of own ad spend in both national and local media. However, the second column indicates that perceived quality also increases with competitors' ad spend in both national and local media. It certainly could be possible for a brand's attitude metrics to increase with competitors' advertising, for example, if competitors' ads draw new consumers to the category who engage in search to discover multiple brands' offerings and attributes. However, within the context of mature brands that advertise regularly, this would seem like an incongruous finding, as most of the brands in this study are already widely known at the start of the sample period. Our expectation prior to conducting this research was that brand attitudes were likely to fall, or at least not increase, with competitors' ad spend.

%A greater surprise is that perceived value both increases {\it and} decreases with lagged advertising in national traditional media. The confusing pattern of both positive/significant {\it and} negative/significant results appears in {\it all} of the six media/metric/ad-type combinations with multiple statistically significant parameter estimates. In summary, such results are quite difficult to interpret and seem to cast doubt on the validity of the findings.

We summarize and interpret the main findings in the all-controls model as follows:
\begin{itemize}
\item Own national traditional ad spend increases all three brand attitude metrics--perceived quality, perceived value and recent satisfaction--with multiple significant lags. 
\item Own local traditional ad spend tends to improve perceived quality and perceived value metrics, but it does not detectably alter recent satisfaction.
\item Own digital advertising increases perceived value, but does not systematically change perceived quality or recent satisfaction.
\item Competitors' national traditional ad spend negatively impacts perceived quality, but does not reliably change perceived value or recent satisfaction metrics.
\item Competitors' local traditional advertising tends to reduce brands' perceived quality, perceived value and recent satisfaction metrics. 
\item Competitors' digital ads tend to reduce perceived quality and recent satisfaction measures. Surprisingly, competitor digital ads seem to increase perceived value.
\end{itemize}

%\textcolor{blue}{As mentioned previously, we interpret these estimates as average treatment effects, which must be lower bounds The reported ad parameter estimates in all-controls model are in order of magnitude of $.0001\%$ or smaller in absolute values. The small effect sizes can be viewed as empirical lower-bounds of average ad effectiveness, mainly because our survey sample is selected independent of ad targeting. If the ads are narrowly or specifically targeted, the survey sample may under-represent the treatment group. If the ads are broadly or randomly targeted, the survey sample may approximate the treatment group. As we are not aware of the exact targeting strategies of over 500 brands in the sample, we are unable to adjust our estimates to obtain the exact treatment effects on treated individuals. However, the reported effect sizes still approximate conservative, lower-bound estimates of ad effectiveness that advertisers can expect from a representative sample.}

%Table \ref{tb:lagdvest} presents the estimates of lagged brand attitudes within each model. As one would expect, brand attitudes are strongly correlated across time and metrics. Inclusion of the brand/quarter control variables primarily changes the correlation between each dependent variable and its own lags from positive to negative. This is because the brand/quarter estimates control for local trends in brand attitudes, so the model including brand/quarter fixed effects produces autocorrelation parameter estimates that reflect the tendency toward local mean reversion observed in Figure \ref{fig:Evid}.
 
To summarize the primary findings, the all-controls model indicates that ({\it i}) brand attitude metrics all rise with multiple lags of the brand's own national traditional advertising; ({\it ii}) local traditional ads increase quality and value perceptions; ({\it iii}) digital ads increase perceived value; ({\it iv}) the effects of competitors' ads are generally negative. Further, inclusion of proper control variables produces patterns of effects that appear more consistent with expectations than descriptive results without controls, without major reductions in the number of statistically significant parameter estimates. 

\subsection{Possible reverse causality: do lagged brand attitudes cause ad spend?}

Based on our understanding of standard advertising practices, we believe that most brands do not adjust their ad spend based on recent changes in weekly brand attitudes. We believe that brands typically set quarterly or annual ad budgets, far in advance, and allocate those budgets to weeks and media vehicles in ways that typically are not driven by recent changes in brand attitude data. However, if that understanding is wrong, then some of the results reported in section \ref{s:findings} may be spurious. 

To investigate, we reversed the all-controls specification, regressing ad spend (in each type of media) on 13 lags of each type of ad spend and five lags of each of the three brand attitude variables. Table \ref{tb:adpred_att} presents the parameter estimates corresponding to lags of brand attitude variables. Out of 45 parameters%(five lags x three brand attitude measures x three types of ad spending)
, only two coefficients (4\%) are statistically significant at the 95\% confidence level, commensurate with expected levels of Type I error. We therefore conclude that simultaneity is not a primary driver of the findings.

\renewcommand{\arraystretch}{1.3}
\begin{table}%[tb]
\caption{Effects of Lagged Attitudes on Contemporaneous Ad Spend}
%\vspace{1em}
\label{tb:adpred_att}
\begin{center}
{
\footnotesize{
\tabcolsep=0.15cm
\begin{tabular}{ll|rl|rl|rl}
\hline															
\multicolumn{2}{l|}{Ad Spend D.V.}			&	\multicolumn{2}{c|}{Nat'l Trad.}			&	\multicolumn{2}{c|}{Loc. Trad.}			&	\multicolumn{2}{c}{Digital}			\\
\hline															
$q_{b,t-\tau}$	&	($\tau=1$)	&	.006	&		&	.074	&		&	.082	&		\\
	&	($\tau=2$)	&	.055	&		&	-.263	&		&	.154	&		\\
	&	($\tau=3$)	&	.741	&		&	.091	&		&	.128	&		\\
	&	($\tau=4$)	&	-.268	&		&	.706	&	*	&	.028	&		\\
	&	($\tau=5$)	&	-.148	&		&	-.171	&		&	.429	&	*	\\
\hline															
$v_{b,t-\tau}$	&	($\tau=1$)	&	.398	&		&	.653	&		&	-.130	&		\\
	&	($\tau=2$)	&	.151	&		&	-.422	&		&	-.055	&		\\
	&	($\tau=3$)	&	.329	&		&	.273	&		&	-.114	&		\\
	&	($\tau=4$)	&	.383	&		&	.536	&		&	.087	&		\\
	&	($\tau=5$)	&	-.041	&		&	-.176	&		&	.337	&		\\
\hline															
$s_{b,t-\tau}$	&	($\tau=1$)	&	-.516	&		&	.269	&		&	.134	&		\\
	&	($\tau=2$)	&	.059	&		&	-.406	&		&	.208	&		\\
	&	($\tau=3$)	&	-.055	&		&	.149	&		&	.420	&		\\
	&	($\tau=4$)	&	.255	&		&	-.550	&		&	.107	&		\\
	&	($\tau=5$)	&	.639	&		&	.005	&		&	-.222	&		\\
\hline															
\multicolumn{2}{l|}{Adj. R Squared}			&	.770	&		&	.740	&		&	.870	&		\\
\hline															
\multicolumn{8}{l}{* Significant at the 95\% confidence level.}
\end{tabular}}}
\end{center}
\end{table}

\subsection{Ad effects by industry}

Seeking a deeper understanding of the drivers of the main results, we re-estimated the all-controls model within industry-specific partitions.\footnote{Industry/week fixed effects were essentially replaced by a separate set of week fixed effects estimated within each partition.} Table \ref{tb:effqua_ind} presents findings from the perceived quality model with statistically significant effects in bold. %Industries are presented in descending order of mean brand attitudes as in Figure \ref{tb:brba}.
Results from the perceived value and recent satisfaction models are presented in the appendix.

The main takeaway is that, despite numerous brands available for each industry, the industry-specific effects exhibit weak statistical power relative to the results calibrated on the full sample. The industry-specific estimates exhibit rates of statistical significance that approximate that expected from Type I error alone. Within national traditional ads, only 16 of 222 estimates (7.2\%) are statistically significant at the 95\% confidence level; comparable figures for local traditional and digital, respectively, are 12 of 222 (5.2\%) and 11 of 222 (5.0\%).

\subsection{All Brand Attitude Metrics}

We restricted primary attention in the analysis to three particular brand attitude variables that we thought were most likely to be influenced by advertising and to matter to advertisers. How do the effects look when we consider all seven available brand attitude metrics? To investigate, we estimated the all-controls model, but this time for each of the seven brand attitude metrics, and including 13 lags of all seven metrics in each of the seven models.

Table \ref{tb:adall7} ad parameter estimates for all seven models. The qualitative conclusions for the three metrics we have focused on (perceived quality, perceived value and recent satisfaction) are nearly identical to the findings reported in section \ref{s:findings}, showing robustness of the estimates to the set of brand attitudes considered. The next three attitude metrics (willingness to recommend, general affect, proud to work) can be positively influenced by own advertising in traditional media; and the proud-to-work attitude is positively related to own digital advertising. Relationships to competitor advertising are mixed.

\renewcommand{\arraystretch}{1.5}
\begin{landscape}
\begin{table}%[tb]
\caption{Ad Parameter Estimates by Industry on Perceived Quality}
%\vspace{1em}
\label{tb:effqua_ind}
\begin{center}
{
\tiny{
\tabcolsep=0.01cm
\begin{tabular}{l|cccccc|cccccc|cccccc}
\hline																																					
	&	\multicolumn{6}{c|}{National Traditonal Ads}											&	\multicolumn{6}{c|}{Local Traditional Ads}											&	\multicolumn{6}{c}{Digital Ads}											\\
\hline																																					
	&	$\tau=0$	&	$\tau=1$	&	$\tau=2$	&	$\tau=3$	&	$\tau=4$	&	$\tau=5$	&	$\tau=0$	&	$\tau=1$	&	$\tau=2$	&	$\tau=3$	&	$\tau=4$	&	$\tau=5$	&	$\tau=0$	&	$\tau=1$	&	$\tau=2$	&	$\tau=3$	&	$\tau=4$	&	$\tau=5$	\\
\hline																																					
Consumer Goods	&	{\bf  4.88e-04}	&	-3.83E-04	&	2.12E-04	&	5.56E-06	&	{\bf  4.47e-04}	&	-2.10E-04	&	2.60E-05	&	{\bf -3.77e-04}	&	2.07E-04	&	-4.25E-05	&	{\bf -3.93e-04}	&	1.89E-04	&	-9.58E-05	&	2.90E-04	&	-3.89E-04	&	2.23E-04	&	-2.80E-04	&	-1.95E-04	\\
Tools/Hardware	&	3.56E-04	&	1.05E-04	&	1.93E-04	&	-3.06E-05	&	-3.69E-04	&	2.48E-04	&	1.26E-04	&	4.32E-05	&	-3.22E-04	&	-9.24E-05	&	-9.45E-05	&	3.72E-04	&	1.11E-03	&	-1.06E-03	&	-7.00E-04	&	8.76E-04	&	1.12E-04	&	2.57E-04	\\
Soft Drinks	&	-3.76E-04	&	1.49E-04	&	{\bf  6.30e-04}	&	6.74E-05	&	{\bf -4.52e-04}	&	3.13E-04	&	2.02E-04	&	1.89E-04	&	-1.84E-04	&	-4.73E-04	&	1.94E-04	&	3.03E-05	&	8.51E-05	&	2.20E-04	&	-1.41E-04	&	4.07E-05	&	5.44E-04	&	-1.89E-05	\\
Beverages: General	&	-1.52E-05	&	-4.35E-06	&	-2.06E-04	&	-7.28E-05	&	5.26E-05	&	{\bf  2.45e-04}	&	-1.05E-04	&	-9.12E-05	&	8.75E-05	&	1.62E-04	&	1.84E-05	&	-5.55E-05	&	2.79E-04	&	8.87E-05	&	3.18E-05	&	-1.56E-04	&	2.15E-05	&	1.97E-04	\\
Media Devices	&	1.56E-05	&	-1.74E-07	&	-5.96E-05	&	1.20E-04	&	-9.73E-05	&	-1.88E-06	&	-6.36E-05	&	1.78E-04	&	2.60E-05	&	1.49E-04	&	{\bf  2.26e-04}	&	-1.50E-04	&	-2.21E-04	&	4.48E-04	&	{\bf -6.83e-04}	&	3.63E-04	&	-2.35E-04	&	3.71E-04	\\
Drugs: OTC	&	-2.91E-04	&	2.14E-04	&	1.52E-04	&	-1.09E-04	&	-2.20E-06	&	-6.79E-05	&	3.10E-04	&	{\bf -3.95e-04}	&	-3.05E-04	&	2.02E-04	&	2.15E-04	&	1.82E-04	&	1.25E-04	&	1.02E-04	&	-2.68E-04	&	-5.96E-05	&	2.21E-05	&	1.64E-05	\\
Electronics:  Audio/Visual	&	-5.77E-05	&	1.94E-04	&	-4.54E-05	&	-1.77E-05	&	{\bf  2.62e-04}	&	-1.79E-04	&	-9.80E-05	&	-1.79E-05	&	-1.33E-04	&	1.78E-06	&	-3.49E-05	&	-1.50E-04	&	{\bf  5.70e-04}	&	8.58E-05	&	-3.64E-04	&	1.64E-04	&	2.07E-04	&	-4.04E-04	\\
Internet Sites	&	-9.34E-05	&	7.77E-05	&	-5.42E-06	&	1.73E-04	&	5.32E-05	&	3.09E-05	&	-1.17E-04	&	7.21E-06	&	-2.25E-04	&	3.28E-06	&	4.90E-05	&	-6.13E-05	&	1.93E-04	&	6.65E-04	&	-9.29E-05	&	2.55E-04	&	3.14E-04	&	2.92E-04	\\
Home/Furnishing Stores	&	1.32E-04	&	-7.43E-05	&	-7.08E-05	&	1.87E-04	&	{\bf  2.64e-04}	&	1.82E-04	&	-5.22E-05	&	1.77E-04	&	8.61E-05	&	4.82E-05	&	1.01E-04	&	1.31E-04	&	4.93E-05	&	-1.20E-05	&	-1.53E-04	&	{\bf  5.89e-04}	&	2.99E-04	&	-2.17E-04	\\
Appliances	&	1.38E-04	&	-2.56E-05	&	2.68E-05	&	6.67E-05	&	-3.06E-05	&	-9.61E-05	&	1.70E-04	&	2.26E-04	&	-1.95E-06	&	-1.01E-04	&	-1.44E-04	&	-6.11E-05	&	4.33E-04	&	{\bf -6.08e-04}	&	{\bf  6.52e-04}	&	7.92E-06	&	4.88E-05	&	5.14E-05	\\
Dept. Stores	&	1.40E-04	&	-4.22E-07	&	{\bf  2.73e-04}	&	-2.58E-06	&	3.13E-05	&	1.93E-04	&	-1.51E-04	&	1.97E-04	&	-5.93E-05	&	1.03E-04	&	1.72E-04	&	2.54E-04	&	-2.15E-05	&	1.08E-04	&	1.52E-04	&	2.05E-04	&	2.60E-05	&	-1.13E-04	\\
Apparel and Shoes	&	2.27E-05	&	1.08E-04	&	8.15E-05	&	1.18E-04	&	1.18E-04	&	-2.96E-05	&	5.10E-05	&	{\bf  2.54e-04}	&	9.12E-05	&	8.02E-05	&	1.04E-05	&	-1.88E-05	&	3.62E-04	&	2.03E-04	&	-2.51E-05	&	1.38E-05	&	-1.65E-04	&	2.67E-04	\\
Car Manufacturers	&	3.11E-04	&	2.15E-05	&	-1.65E-04	&	1.93E-04	&	2.06E-04	&	2.23E-04	&	-1.14E-04	&	-9.55E-05	&	-2.09E-04	&	-8.67E-05	&	1.94E-04	&	2.49E-05	&	2.44E-04	&	3.29E-04	&	9.18E-04	&	-4.03E-04	&	2.22E-04	&	-1.51E-04	\\
TV Networks	&	3.71E-05	&	7.08E-05	&	1.61E-04	&	-4.54E-05	&	-2.16E-05	&	{\bf  3.05e-04}	&	-8.16E-05	&	2.14E-04	&	-1.47E-04	&	2.58E-04	&	-1.73E-04	&	-1.72E-04	&	-8.77E-05	&	-2.06E-05	&	-2.07E-04	&	-1.36E-06	&	-5.30E-05	&	2.18E-05	\\
Hotels	&	-4.80E-05	&	-5.86E-05	&	1.32E-04	&	1.48E-05	&	1.39E-04	&	7.78E-05	&	-8.19E-05	&	-2.78E-05	&	9.55E-05	&	-2.14E-04	&	-2.30E-04	&	-5.03E-05	&	1.26E-04	&	4.21E-05	&	-2.33E-04	&	6.80E-05	&	-4.91E-05	&	9.93E-05	\\
Fast Food	&	-7.06E-05	&	-1.57E-04	&	-1.60E-04	&	1.32E-04	&	-1.60E-04	&	-1.36E-04	&	-5.50E-05	&	-6.49E-05	&	-4.36E-05	&	1.31E-04	&	7.96E-05	&	1.17E-04	&	-2.04E-04	&	-3.31E-05	&	-1.87E-04	&	-1.56E-04	&	-1.16E-04	&	-1.34E-04	\\
Liquor	&	1.27E-04	&	-1.39E-04	&	-1.15E-04	&	-1.33E-04	&	1.44E-04	&	-4.24E-05	&	-1.74E-04	&	8.02E-05	&	6.38E-06	&	-7.82E-05	&	-6.04E-05	&	-6.67E-05	&	2.42E-04	&	-4.33E-05	&	-2.92E-04	&	3.64E-04	&	-2.77E-04	&	1.43E-05	\\
Ice Cream/Pizza/Coffee	&	-1.06E-04	&	{\bf  4.65e-04}	&	-6.42E-05	&	-3.18E-04	&	-5.71E-07	&	5.84E-05	&	9.64E-05	&	-6.59E-05	&	-3.22E-04	&	2.34E-05	&	1.99E-04	&	1.98E-04	&	-3.10E-04	&	4.92E-04	&	-2.22E-04	&	2.48E-04	&	-2.91E-04	&	-2.77E-05	\\
Clothing Stores	&	-1.27E-04	&	-1.40E-04	&	-2.32E-04	&	-2.32E-05	&	-4.52E-06	&	7.12E-05	&	1.19E-04	&	-1.14E-05	&	2.26E-05	&	1.18E-04	&	2.91E-04	&	-3.60E-04	&	-5.40E-04	&	1.01E-04	&	3.29E-04	&	-2.13E-04	&	-3.40E-04	&	2.86E-05	\\
Books/Kids/Office Stores	&	1.32E-04	&	4.02E-05	&	1.27E-04	&	1.07E-04	&	8.13E-06	&	2.05E-04	&	1.22E-04	&	1.04E-04	&	1.10E-04	&	5.72E-05	&	2.16E-04	&	1.89E-04	&	-1.97E-04	&	2.01E-05	&	-2.40E-04	&	2.50E-04	&	-1.31E-05	&	-1.33E-04	\\
Casual Dining	&	5.59E-05	&	1.88E-04	&	1.87E-04	&	-3.89E-05	&	-9.49E-05	&	5.83E-05	&	7.06E-05	&	2.53E-04	&	-1.67E-04	&	-5.52E-05	&	-8.25E-05	&	1.50E-04	&	{\bf -3.89e-04}	&	2.57E-04	&	4.86E-05	&	5.87E-05	&	-5.72E-05	&	-3.29E-04	\\
Beer	&	1.03E-04	&	2.70E-04	&	2.96E-04	&	9.92E-05	&	1.96E-04	&	1.14E-04	&	2.87E-04	&	{\bf  4.58e-04}	&	1.46E-04	&	2.16E-05	&	2.87E-04	&	-3.01E-04	&	{\bf  5.08e-04}	&	{\bf -6.21e-04}	&	-1.13E-04	&	1.61E-04	&	1.16E-04	&	-6.85E-05	\\
Fast Casual Dining	&	-2.92E-06	&	{\bf  3.96e-04}	&	-6.29E-05	&	2.57E-04	&	-8.82E-05	&	-2.35E-04	&	{\bf  2.66e-04}	&	-1.83E-04	&	-5.90E-06	&	1.09E-04	&	-1.51E-04	&	4.62E-05	&	-2.66E-04	&	-1.80E-04	&	{\bf  5.71e-04}	&	-1.63E-04	&	1.68E-04	&	1.07E-04	\\
Gasoline/AutoAccessories	&	-1.44E-04	&	1.70E-04	&	3.84E-05	&	-1.00E-04	&	1.78E-04	&	{\bf -2.29e-04}	&	1.63E-04	&	3.15E-05	&	1.04E-04	&	-1.56E-04	&	-2.16E-05	&	1.06E-05	&	-1.13E-04	&	-9.15E-05	&	2.19E-05	&	-2.89E-04	&	2.28E-04	&	-7.34E-05	\\
Sports/Electronics Stores	&	-1.98E-04	&	4.06E-05	&	-5.57E-05	&	2.13E-04	&	-1.68E-04	&	1.92E-04	&	-2.34E-04	&	-3.56E-05	&	3.33E-05	&	-2.70E-05	&	-1.17E-04	&	2.67E-05	&	9.10E-05	&	-5.25E-05	&	1.38E-04	&	1.19E-04	&	-1.75E-04	&	-3.08E-05	\\
Cruise/TravelAgents	&	1.95E-04	&	1.37E-04	&	1.45E-04	&	{\bf  3.16e-04}	&	2.11E-04	&	-1.38E-04	&	1.14E-04	&	-4.42E-05	&	-1.47E-04	&	1.76E-04	&	-1.03E-04	&	3.23E-04	&	{\bf -6.20e-04}	&	2.47E-04	&	3.37E-04	&	1.14E-04	&	5.38E-05	&	-1.54E-04	\\
Media Services	&	-1.18E-04	&	{\bf  4.59e-04}	&	7.88E-05	&	-6.30E-05	&	2.03E-04	&	-2.26E-05	&	-7.42E-05	&	1.48E-04	&	-8.74E-06	&	1.78E-04	&	8.27E-05	&	8.67E-06	&	2.23E-04	&	-3.48E-05	&	4.75E-04	&	-5.36E-04	&	3.11E-04	&	-3.39E-04	\\
Insurance	&	-1.47E-04	&	8.85E-05	&	9.84E-05	&	8.75E-05	&	-2.14E-05	&	2.16E-04	&	-5.30E-05	&	-2.13E-04	&	3.06E-04	&	1.64E-05	&	5.24E-05	&	1.18E-04	&	1.54E-04	&	-3.41E-05	&	-4.26E-04	&	4.38E-04	&	-5.38E-04	&	3.70E-04	\\
Steakhouses/CasualDining	&	7.08E-05	&	-1.61E-04	&	7.42E-05	&	-7.88E-05	&	1.01E-04	&	-6.54E-05	&	-3.48E-05	&	1.21E-04	&	2.46E-05	&	3.50E-05	&	{\bf -3.00e-04}	&	9.79E-05	&	6.66E-05	&	-7.75E-05	&	-1.31E-04	&	1.56E-04	&	7.44E-05	&	2.43E-04	\\
Women's Clothing Stores	&	2.27E-04	&	2.69E-04	&	2.61E-04	&	9.68E-05	&	1.58E-04	&	2.81E-04	&	3.23E-05	&	-9.97E-05	&	-8.01E-05	&	{\bf -4.13e-04}	&	-2.14E-04	&	6.92E-05	&	3.52E-04	&	-2.45E-04	&	-1.05E-04	&	-2.41E-04	&	{\bf  6.66e-04}	&	6.34E-05	\\
Airlines	&	9.96E-05	&	1.52E-04	&	-5.11E-05	&	-2.18E-05	&	-4.36E-05	&	-7.93E-05	&	-1.86E-05	&	-8.77E-05	&	4.57E-05	&	{\bf  2.21e-04}	&	1.19E-05	&	{\bf  2.48e-04}	&	-4.48E-05	&	1.92E-04	&	-1.30E-04	&	-3.32E-04	&	1.66E-04	&	5.67E-05	\\
Casinos	&	4.90E-05	&	-1.22E-04	&	-1.74E-04	&	-1.21E-04	&	3.70E-06	&	-5.01E-05	&	1.62E-05	&	-1.68E-04	&	-3.85E-05	&	1.02E-04	&	1.53E-04	&	-2.34E-05	&	4.80E-05	&	4.61E-04	&	-3.37E-04	&	5.62E-04	&	4.22E-04	&	4.41E-04	\\
Financial Services	&	4.82E-05	&	1.16E-04	&	-3.30E-05	&	1.19E-04	&	{\bf  2.30e-04}	&	1.08E-04	&	1.02E-04	&	1.95E-04	&	{\bf -2.25e-04}	&	5.26E-05	&	8.55E-05	&	-1.44E-04	&	2.60E-05	&	9.77E-06	&	1.72E-04	&	-4.26E-04	&	-2.17E-04	&	1.33E-04	\\
Grocery Stores	&	5.35E-05	&	-2.30E-04	&	{\bf  3.91e-04}	&	-2.18E-04	&	-1.28E-04	&	7.46E-05	&	-4.04E-04	&	5.07E-05	&	-3.97E-04	&	9.17E-05	&	-9.33E-05	&	-3.20E-05	&	1.43E-04	&	-6.58E-05	&	-1.42E-04	&	-1.58E-05	&	-3.95E-05	&	1.12E-04	\\
Drugs: General	&	1.66E-04	&	-1.46E-04	&	-1.32E-04	&	-9.10E-05	&	1.65E-05	&	1.37E-04	&	2.12E-05	&	-1.50E-05	&	1.35E-04	&	8.29E-05	&	-5.35E-05	&	-9.88E-06	&	-5.85E-05	&	-5.37E-04	&	3.40E-04	&	7.74E-05	&	-4.12E-05	&	2.42E-04	\\
Drugs: Prescription	&	-2.99E-05	&	3.06E-05	&	-2.45E-04	&	1.38E-04	&	-8.10E-05	&	1.59E-04	&	3.34E-04	&	3.07E-05	&	-9.73E-05	&	1.52E-04	&	4.65E-05	&	2.92E-04	&	2.06E-04	&	1.63E-04	&	-4.09E-04	&	2.02E-04	&	-4.24E-04	&	4.36E-05	\\
Consumer Banks	&	-2.26E-04	&	-7.01E-05	&	1.10E-04	&	-1.45E-04	&	-1.82E-04	&	-8.28E-05	&	8.38E-05	&	-4.08E-05	&	-3.79E-05	&	9.34E-05	&	-5.13E-05	&	-2.80E-05	&	-2.91E-05	&	2.66E-04	&	5.96E-05	&	1.47E-05	&	-1.82E-04	&	-1.16E-05	\\
\hline																																							
\multicolumn{19}{l}{Note: Estimates in bold are statistically significant at the 95\% level.}
\end{tabular}}}
\end{center}
\end{table}
\end{landscape}

\renewcommand{\arraystretch}{1}
\begin{table}%[tb]
\caption{Ad Parameter Estimates for All Brand Attitude D.V.'s}
%\vspace{1em}
\label{tb:adall7}
\begin{center}
{
\scriptsize{
\tabcolsep=0.05cm
\begin{tabular}{ll|rlrlrlrlrlrlrl}
\hline																															
	\multicolumn{2}{l|}{Brand Attitude}		&	\multicolumn{2}{c}{Perceived}			&	\multicolumn{2}{c}{Perceived}			&	\multicolumn{2}{c}{Recent}			&	\multicolumn{2}{c}{Willing to}			&	\multicolumn{2}{c}{General}			&	\multicolumn{2}{c}{Proud to}			&	\multicolumn{2}{c}{Heard}			\\
	\multicolumn{2}{l|}{D.V.}		&	\multicolumn{2}{c}{Quality}			&	\multicolumn{2}{c}{Value}			&	\multicolumn{2}{c}{Satisfaction}			&	\multicolumn{2}{c}{Recommend}			&	\multicolumn{2}{c}{Affect}			&	\multicolumn{2}{c}{Work}			&	\multicolumn{2}{c}{About}			\\
\hline																															
Own	&	($\tau=0$)	&	3.06E-05	&		&	3.89E-05	&		&	2.35E-05	&		&	7.90E-06	&		&	2.11E-05	&		&	3.37E-05	&		&	1.32E-04	&	**	\\
Nat'l	&	($\tau=1$)	&	6.39E-05	&	**	&	7.20E-05	&	**	&	1.59E-05	&		&	6.40E-05	&	**	&	5.50E-05	&	*	&	6.03E-05	&	**	&	2.55E-04	&	**	\\
Trad.	&	($\tau=2$)	&	1.49E-05	&		&	5.19E-05	&	*	&	3.48E-05	&	*	&	6.63E-06	&		&	8.24E-06	&		&	-5.91E-06	&		&	1.23E-04	&	**	\\
Ads	&	($\tau=3$)	&	1.93E-05	&		&	7.84E-06	&		&	3.76E-05	&	*	&	3.55E-05	&		&	1.74E-05	&		&	4.47E-05	&	*	&	9.23E-05	&	**	\\
	&	($\tau=4$)	&	5.32E-05	&	*	&	1.80E-05	&		&	3.26E-05	&	*	&	5.94E-06	&		&	-5.74E-06	&		&	1.24E-05	&		&	1.00E-04	&	**	\\
	&	($\tau=5$)	&	2.67E-05	&		&	-2.94E-05	&		&	2.44E-06	&		&	5.57E-06	&		&	2.22E-06	&		&	1.96E-05	&		&	1.12E-04	&	**	\\
\hline																															
Own	&	($\tau=0$)	&	2.91E-05	&		&	3.17E-05	&		&	-1.87E-05	&		&	5.21E-05	&	*	&	3.85E-05	&		&	3.48E-05	&		&	1.29E-04	&	**	\\
Local	&	($\tau=1$)	&	6.24E-05	&	**	&	3.29E-05	&		&	1.22E-05	&		&	3.68E-05	&		&	6.67E-05	&	**	&	8.39E-05	&	**	&	1.36E-04	&	**	\\
Trad.	&	($\tau=2$)	&	-1.85E-05	&		&	1.60E-05	&		&	-2.60E-05	&		&	1.97E-05	&		&	6.37E-08	&		&	3.00E-05	&		&	1.03E-04	&	**	\\
Ads	&	($\tau=3$)	&	2.64E-05	&		&	1.11E-06	&		&	-1.22E-05	&		&	-6.13E-06	&		&	4.66E-05	&		&	3.37E-05	&		&	1.10E-04	&	**	\\
	&	($\tau=4$)	&	1.41E-05	&		&	-2.59E-06	&		&	-1.02E-05	&		&	2.68E-05	&		&	5.78E-06	&		&	2.25E-05	&		&	5.95E-05	&	**	\\
	&	($\tau=5$)	&	2.33E-05	&		&	6.45E-05	&	**	&	8.77E-06	&		&	1.39E-05	&		&	3.62E-05	&		&	-2.28E-05	&		&	9.75E-05	&	**	\\
\hline																															
Own 	&	($\tau=0$)	&	2.54E-05	&		&	2.40E-05	&		&	1.37E-06	&		&	5.22E-05	&		&	9.35E-06	&		&	1.20E-04	&	**	&	5.80E-05	&		\\
Digital	&	($\tau=1$)	&	5.63E-05	&		&	-5.93E-05	&		&	2.62E-05	&		&	-6.81E-05	&		&	4.94E-05	&		&	-2.25E-05	&		&	-2.24E-05	&		\\
Ads	&	($\tau=2$)	&	-4.26E-05	&		&	7.85E-05	&		&	-2.94E-06	&		&	6.13E-05	&		&	2.83E-05	&		&	2.11E-05	&		&	4.68E-05	&		\\
	&	($\tau=3$)	&	2.83E-05	&		&	-6.03E-05	&		&	-3.02E-06	&		&	-1.66E-06	&		&	8.22E-07	&		&	-1.13E-05	&		&	4.38E-05	&		\\
	&	($\tau=4$)	&	-1.09E-05	&		&	2.29E-05	&		&	-1.03E-05	&		&	1.72E-05	&		&	4.28E-05	&		&	-1.19E-05	&		&	1.93E-05	&		\\
	&	($\tau=5$)	&	-1.23E-05	&		&	2.81E-05	&		&	-8.79E-06	&		&	5.10E-06	&		&	-2.87E-05	&		&	1.87E-05	&		&	4.27E-05	&		\\
\hline																															
Comp.	&	($\tau=0$)	&	-4.46E-04	&	*	&	1.06E-04	&		&	-1.09E-04	&		&	1.28E-05	&		&	1.43E-04	&		&	8.59E-05	&		&	-1.80E-04	&		\\
Nat'l	&	($\tau=1$)	&	-4.36E-04	&	*	&	-3.87E-04	&		&	6.75E-05	&		&	-5.84E-04	&	**	&	-2.58E-04	&		&	2.51E-05	&		&	-8.95E-04	&	**	\\
Trad.	&	($\tau=2$)	&	-3.42E-04	&		&	-1.70E-04	&		&	-1.42E-05	&		&	-1.51E-04	&		&	-3.30E-04	&		&	1.54E-04	&		&	-1.87E-04	&		\\
Ads	&	($\tau=3$)	&	-4.95E-04	&	*	&	-3.00E-04	&		&	-1.64E-05	&		&	2.92E-04	&		&	2.00E-05	&		&	5.17E-04	&	*	&	-6.53E-04	&	**	\\
	&	($\tau=4$)	&	-9.49E-05	&		&	-4.91E-05	&		&	9.86E-05	&		&	-2.92E-04	&		&	7.19E-05	&		&	-1.40E-04	&		&	1.60E-04	&		\\
	&	($\tau=5$)	&	4.92E-05	&		&	-3.33E-04	&		&	-4.89E-05	&		&	-2.52E-04	&		&	9.01E-05	&		&	-2.22E-04	&		&	-1.89E-04	&		\\
\hline																															
Comp.	&	($\tau=0$)	&	1.37E-04	&		&	-3.25E-04	&		&	-3.23E-04	&	*	&	-3.99E-04	&		&	-1.05E-04	&		&	3.05E-04	&		&	-2.40E-05	&		\\
Local	&	($\tau=1$)	&	2.30E-04	&		&	-4.09E-04	&		&	-2.62E-04	&		&	1.73E-04	&		&	3.42E-04	&		&	-3.80E-04	&		&	5.03E-06	&		\\
Trad.	&	($\tau=2$)	&	-5.25E-04	&	*	&	4.97E-05	&		&	-6.31E-05	&		&	1.29E-05	&		&	6.63E-05	&		&	-6.73E-05	&		&	4.85E-05	&		\\
Ads	&	($\tau=3$)	&	2.04E-04	&		&	-2.17E-04	&		&	2.28E-04	&		&	-1.06E-04	&		&	4.44E-04	&	*	&	-1.55E-05	&		&	9.24E-05	&		\\
	&	($\tau=4$)	&	-8.36E-05	&		&	6.34E-05	&		&	-1.01E-04	&		&	1.10E-04	&		&	3.22E-05	&		&	-8.14E-06	&		&	-3.09E-04	&		\\
	&	($\tau=5$)	&	-3.38E-04	&		&	1.82E-04	&		&	-1.03E-04	&		&	9.54E-05	&		&	1.08E-04	&		&	2.62E-04	&		&	-1.00E-04	&		\\
\hline																															
Comp. 	&	($\tau=0$)	&	1.59E-04	&		&	-5.74E-04	&		&	6.02E-05	&		&	6.28E-04	&	*	&	-1.06E-03	&	**	&	1.90E-04	&		&	-6.59E-04	&	*	\\
Digital	&	($\tau=1$)	&	5.16E-04	&		&	-3.56E-04	&		&	-2.19E-04	&		&	-5.35E-04	&		&	1.03E-03	&	**	&	-2.13E-04	&		&	-5.55E-04	&		\\
Ads	&	($\tau=2$)	&	-2.47E-04	&		&	4.11E-04	&		&	4.33E-04	&		&	1.01E-04	&		&	6.02E-04	&		&	-4.38E-04	&		&	-4.63E-04	&		\\
	&	($\tau=3$)	&	1.40E-04	&		&	7.19E-05	&		&	-3.33E-04	&		&	4.05E-04	&		&	-2.56E-04	&		&	-3.25E-04	&		&	1.14E-04	&		\\
	&	($\tau=4$)	&	-1.12E-04	&		&	-1.32E-04	&		&	-2.58E-04	&		&	-1.54E-04	&		&	6.82E-04	&	*	&	2.74E-04	&		&	8.20E-05	&		\\
	&	($\tau=5$)	&	-7.82E-04	&	*	&	7.91E-04	&	**	&	-7.97E-04	&	**	&	3.64E-04	&		&	-4.80E-04	&		&	1.62E-04	&		&	-2.40E-04	&		\\
\hline																															
\multicolumn{2}{l|}{Adj. R Squared}			&	.971	&		&	.966	&		&	.983	&		&	.968	&		&	.972	&		&	.961	&		&	.947	&		\\
\hline																											\multicolumn{16}{l}{* Significant at the 95\% confidence level.} \\
\multicolumn{16}{l}{** Significant at the 99\% confidence level.} 
\end{tabular}}}
\end{center}
\end{table}

The final attitude metric (heard about) is extremely strongly related to all lags of own advertising in traditional media. These results are unsurprising; the survey instrument explicitly asks about recent advertising exposure. What {\it is} surprising is the absence of any detectable relationship between ``heard about'' and own digital advertising. However, it is the case that ``heard about'' decreases with contemporaneous competitor digital advertising. Recall that YouGov panelists answer brand attitude questions online, so the results came from respondents who may even be skewed more toward digital advertising exposures than the overall population.%We do not necessarily conclude that this means that digital advertising {\it cannot} be used for branding purposes; it may instead mean that digital advertising {\it typically was not} used for branding purposes in this sample period; %\footnote{\textcolor{red}{Google has recently started to promote their digital advertising for brand building and argued that digital media have not been a brand manager's toolkit. For more details, see https://www.thinkwithgoogle.com/advertising-channels/search/how-advertisers-are-using-search-for-brand-building/.}}
%or that the effects are dampened by measurement error in the measures of digital advertising; or any combination thereof. We find this to be a compelling question for future research. 

\subsection{Temporal aggregation}

A frequent question in the advertising literature is how temporal (dis)aggregation affects estimation results (see, e.g., \citealt{TF:2006} and references therein). To investigate, we aggregated the brand attitude and ad spend variables into two-week and four-week intervals, then ran comparable versions of the all-controls specification within each dataset. The qualitative results using two-week interval data, which are provided in Table \ref{tb:2wkpar} in the appendix, are very similar to weekly-level all-controls model. The four-week results in Table \ref{tb:4wkpar} are also quite similar, though less so. The aggregated data yield higher proportions of advertising parameters that are statistically significant at the 95\% level, with 39\% of advertising parameters exhibiting statistical significance in the four-week data, followed by 31\% and 20\% in the two-week and one-week datasets, respectively. We favor the results based on weekly data, as we believe that they are more conservative and that the weekly data enable better controls for unobserved confounds.

\section{Conclusions, Limitations and Implications}
%\textcolor{red}{[For Wes's last point, I think Wes misunderstood the goal of the third paragraph. My understanding is that Wes thinks that the substantive implication on digital ads stretches our results too far. So, I tried to clarify that the paragraph is to discuss that we cannot distinguish whether the low effect of digital ads is a statistical problem or the actual null effect. Feel free to edit my 3rd paragraph, if you disagree. I actually prefer to and suggest to delete the entire 3rd paragraph, as I don't see a reason to explain it. The original 3rd paragraph is hidden as annotation next to my edits.]}

In this research, we analyzed a unique ``large-N, large-T'' panel dataset of brand attitudes and advertising expenditures to investigate three specific research questions. We applied straightforward models to comparable metrics to investigate how ads in different media may change consumers' brand attitudes, subject to a clear identifying assumption. We further showed how those effects are impacted by various controls for unobserved variables, finding that industry/week and brand/quarter fixed effects are individually and jointly important control variables whose inclusion brings advertising parameter estimates closer to expectations without major reductions in estimation precision. Although the controls employed may not apply perfectly to every brand in the sample, we believe that the identifying assumption is reasonable for most of the brands considered, and that the overall results are likely to approximate the true effects. 

The primary learnings indicate that ({\it i}) brand attitude metrics all rise with multiple lags of the brand's own national traditional advertising; ({\it ii}) local traditional ads increase perceived quality and perceived value; ({\it iii}) digital ads increase perceived value; and ({\it iv}) the effects of competitors' ads are generally negative. The qualitative results are robust, as the data indicate that they are not solely driven by the set of brand attitudes considered, the number of lags of ad spend included in the model, the assumption that ad spend precedes brand attitudes, or the temporal disaggregation of the data. %The data indicate that, consistent with our understanding of common industry practices, reverse causality is not an important concern: ad spend is not driven by recent changes in brand attitudes.

We hope that the findings and control strategies offered may aid marketers and their agencies in using data to guide important practical questions such as whether to advertise, how much to spend, and how to allocate ad budgets. Such empirical guidance may be especially needed in industries where available data complicate the estimation of causal effects of ads on sales, such as markets with long purchase cycles or long inter-purchase times. However, it is important to note that the concerns about statistical power that have been raised in the advertising/sales literature also apply to advertising effects on brand attitudes. Practitioners interested in estimating precise effects of ad spend on brand attitudes should seriously consider running digital experiments, randomizing traditional ad spend across time and geography, and using the quasi-experimental research designs outlined in section 3. The brand attitude data analyzed in this paper provide insufficient power to estimate precise industry-specific effects, so brand-specific effects would be even more difficult to estimate. We therefore advise investigation of other intermediate metrics as candidate advertising response measures, such as store traffic, consideration, or information acquisition via online search (e.g. Du et al. 2017). %It seems unlikely that many brands can use ``media mix modeling'' to estimate precise ad effects on brand attitudes using their own data alone, so practitioners should exercise appropriate caution and skepticism when considering vendors that offer such services. %using regional variation along with intertemporal variation; by, for example, contracting with a provider of brand attitude data that reports regional differences in , and ideally one of the quasi-experimental strategies outlined in section 3, as the industry-specific regressions reported in this paper did not produce precise estimates.

This research is subject to numerous caveats and limitations. Prominent among them is that the findings and control strategies only apply to the set of brands studied, i.e. mature brands that advertise regularly. We believe they will be of limited use in evolving categories, for new brands, or for brands that advertise irregularly. Understanding the links between ad spend and brand attitudes in those situations therefore remains as another topic for future research, as it likely requires customized approaches to control for unobserved variables that drive both firm advertising and brand attitudes.

We believe the most important implication of these findings is a renewed call for highly powered field experiments. Ideally these would run simultaneously across multiple types of media, vary treatments across time and space, allow for interactions between media, and estimate treatment effects on multiple comparable behavioral and attitudinal metrics. In particular, such ambitions should become increasingly feasible as more TV advertising is delivered digitally and additional targeting capabilities are brought to market (\citealt{TNG:forth}). We believe the advertising industry will eventually reach the point that scientific understanding of causal ad effects is used to set media budgets that can be provably linked to profit-relevant outcomes. We hope the results in this paper will offer a useful signpost to help guide hypotheses and statistical power calculations as the industry makes progress toward such efforts.

\newpage
\bibliographystyle{ormsv080}  % amsplain, ormsv080
\bibliography{AdBrand}

\newpage
\subsection*{Appendix}
This appendix presents information and results that are not included in the main body for brevity. Table \ref{tb:brind} lists all brands in the sample by industry. Tables \ref{tb:est1}-\ref{tb:est4} present ad parameter estimates and their standard errors in descriptive models, models with industry/week controls, models with brand/quarter controls, and all-controls models, respectively. Tables \ref{tb:lagdv1}-\ref{tb:lagdv4} present parameter estimates and standard errors for lagged dependent variables in all four models. Table \ref{tb:diff_lagsqua} presents ad parameter estimate variation with number of lags included in the perceived quality all-controls model specification. Tables \ref{tb:effval_ind} and \ref{tb:effsat_ind} report industry-specific ad parameters perceived value and recent satisfaction models. Tables \ref{tb:2wkpar} and \ref{tb:4wkpar} indicate results for the all-controls models estimated in data aggregated into two-week and four-week intervals.

\renewcommand{\arraystretch}{1}
\begin{table}[h]
\caption{Summary of Brands in YouGov Data by Industry}
%\vspace{1em}
\label{tb:brind}
\begin{center}
{
\tiny{
\tabcolsep=0.15cm
\begin{tabular}{l|l}
	\hline				
	{\bf Industry}	&		{\bf All Brands}	\\
	\hline				
\cellcolor[gray]{0.9}Consumer Goods	&	\cellcolor[gray]{0.9}Betty Crocker, Campbell's, Charmin, Clorox, Colgate, Crest, Dawn, Gillette, M\&M's, Nabisco, 	\\
\cellcolor[gray]{0.9}	&	\cellcolor[gray]{0.9}Pillsbury, Quaker, Schick, Snickers, Tide	\\
	Tools/Hardware	&		Bosch, Craftsman, DeWalt, Lincoln Electric	\\
\cellcolor[gray]{0.9}Soft Drinks	&	\cellcolor[gray]{0.9}7UP, A \& W, Coca Cola, Dr. Pepper, Mountain Dew, Pepsi, Red Bull, Sprite	\\
	Beverages: General	&		Crystal Light, Dasani, Folgers, Gatorade, Green Mountain, Lipton, Maxwell House, 	\\
		&		Minute Maid, Motts, Powerade, Snapple, Tropicana, V8	\\
\cellcolor[gray]{0.9}Media Devices	&	\cellcolor[gray]{0.9}Acer, Apple, BlackBerry, Brother, Canon, Dell, Gateway, HP, IBM, Intel, LG, Motorola, 	\\
\hline				
\end{tabular}}}
\end{center}
\end{table}

\renewcommand{\arraystretch}{1}
\begin{table}[h]
\caption{Ad Parameter Estimates in Descriptive Model}
%\vspace{1em}
\label{tb:est1}
\begin{center}
{
\scriptsize{
\tabcolsep=0.15cm
\begin{tabular}{ll|ccl|ccl|ccl}
\hline																					
\multicolumn{2}{l|}{Brand Attitude D.V.}			&	\multicolumn{3}{c|}{Perceived Quality}					&	\multicolumn{3}{c|}{Perceived Value}					&	\multicolumn{3}{c}{Recent Satisfaction}					\\
\hline																					
\multicolumn{2}{l|}{Ad Spend}			&	Est.	&	Std.Err.	&		&	Est.	&	Std.Err.	&		&	Est.	&	Std.Err.	&		\\
\hline																					
Own National	&	($\tau=0$)	&	4.12E-05	&	2.15E-05	&		&	4.00E-05	&	2.04E-05	&	*	&	2.72E-05	&	1.54E-05	&		\\
Trad. Ads	&	($\tau=1$)	&	1.04E-04	&	2.26E-05	&	**	&	6.71E-05	&	2.14E-05	&	**	&	2.37E-06	&	1.62E-05	&		\\
	&	($\tau=2$)	&	-2.12E-05	&	2.26E-05	&		&	5.80E-05	&	2.14E-05	&	**	&	2.58E-05	&	1.62E-05	&		\\
	&	($\tau=3$)	&	2.63E-05	&	2.26E-05	&		&	4.96E-06	&	2.14E-05	&		&	2.03E-05	&	1.62E-05	&		\\
	&	($\tau=4$)	&	4.13E-05	&	2.26E-05	&		&	-1.71E-05	&	2.14E-05	&		&	1.79E-05	&	1.62E-05	&		\\
	&	($\tau=5$)	&	-2.60E-05	&	2.15E-05	&		&	-5.56E-05	&	2.04E-05	&	**	&	-2.32E-05	&	1.54E-05	&		\\
\end{tabular}}}
\end{center}
\end{table}

\renewcommand{\arraystretch}{1}
\begin{table}%[tb]
\caption{Ad Parameter Estimates in Model with Industry/Week Control}
%\vspace{1em}
\label{tb:est2}
\begin{center}
{
\scriptsize{
\tabcolsep=0.15cm
\begin{tabular}{ll|ccl|ccl|ccl}
\hline																					
\multicolumn{2}{l|}{Brand Attitude D.V.}			&	\multicolumn{3}{c|}{Perceived Quality}					&	\multicolumn{3}{c|}{Perceived Value}					&	\multicolumn{3}{c}{Recent Satisfaction}					\\
\hline																					
\multicolumn{2}{l|}{Ad Spend}			&	Est.	&	Std.Err.	&		&	Est.	&	Std.Err.	&		&	Est.	&	Std.Err.	&		\\
\hline																				
Own National	&	($\tau=0$)	&	4.00E-05	&	2.05E-05	&		&	2.63E-05	&	1.98E-05	&		&	2.77E-05	&	1.50E-05	&		\\
Trad. Ads	&	($\tau=1$)	&	6.51E-05	&	2.15E-05	&	**	&	6.15E-05	&	2.08E-05	&	**	&	9.40E-06	&	1.58E-05	&		\\
	&	($\tau=2$)	&	3.57E-06	&	2.15E-05	&		&	4.15E-05	&	2.08E-05	&	*	&	2.51E-05	&	1.58E-05	&		\\
	&	($\tau=3$)	&	1.24E-05	&	2.15E-05	&		&	-9.58E-06	&	2.08E-05	&		&	1.86E-05	&	1.58E-05	&		\\
	&	($\tau=4$)	&	4.65E-05	&	2.15E-05	&	*	&	9.09E-07	&	2.08E-05	&		&	9.86E-06	&	1.58E-05	&		\\
	&	($\tau=5$)	&	-1.29E-06	&	2.05E-05	&		&	-4.90E-05	&	1.98E-05	&	*	&	-2.53E-05	&	1.50E-05	&		\\
\end{tabular}}}
\end{center}
\end{table}

\renewcommand{\arraystretch}{1}
\begin{table}%[tb]
\caption{Ad Parameter Estimates in Model with Brand/Quarter Control}
%\vspace{1em}
\label{tb:est3}
\begin{center}
{
\scriptsize{
\tabcolsep=0.15cm
\begin{tabular}{ll|ccl|ccl|ccl}
\hline																					
\multicolumn{2}{l|}{Brand Attitude D.V.}			&	\multicolumn{3}{c|}{Perceived Quality}					&	\multicolumn{3}{c|}{Perceived Value}					&	\multicolumn{3}{c}{Recent Satisfaction}					\\
\hline																					
\multicolumn{2}{l|}{Ad Spend}			&	Est.	&	Std.Err.	&		&	Est.	&	Std.Err.	&		&	Est.	&	Std.Err.	&		\\
\hline																				
Own National	&	($\tau=0$)	&	2.99E-05	&	2.22E-05	&		&	4.72E-05	&	2.11E-05	&	*	&	2.06E-05	&	1.59E-05	&		\\
Trad. Ads	&	($\tau=1$)	&	1.16E-04	&	2.25E-05	&	**	&	8.38E-05	&	2.14E-05	&	**	&	1.30E-05	&	1.61E-05	&		\\
	&	($\tau=2$)	&	1.51E-05	&	2.24E-05	&		&	8.23E-05	&	2.13E-05	&	**	&	4.04E-05	&	1.60E-05	&	*	\\
	&	($\tau=3$)	&	5.39E-05	&	2.23E-05	&	*	&	3.73E-05	&	2.12E-05	&		&	4.13E-05	&	1.60E-05	&	**	\\
	&	($\tau=4$)	&	7.88E-05	&	2.24E-05	&	**	&	2.31E-05	&	2.13E-05	&		&	4.23E-05	&	1.61E-05	&	**	\\
	&	($\tau=5$)	&	4.08E-05	&	2.21E-05	&		&	-1.25E-05	&	2.10E-05	&		&	1.22E-05	&	1.58E-05	&		\\
\end{tabular}}}
\end{center}
\end{table}

\renewcommand{\arraystretch}{1}
\begin{table}%[tb]
\caption{Ad Parameter Estimates in All Controls Model}
%\vspace{1em}
\label{tb:est4}
\begin{center}
{
\scriptsize{
\tabcolsep=0.15cm
\begin{tabular}{ll|ccl|ccl|ccl}
\hline																					
\multicolumn{2}{l|}{Brand Attitude D.V.}			&	\multicolumn{3}{c|}{Perceived Quality}					&	\multicolumn{3}{c|}{Perceived Value}					&	\multicolumn{3}{c}{Recent Satisfaction}					\\
\hline																					
\multicolumn{2}{l|}{Ad Spend}			&	Est.	&	Std.Err.	&		&	Est.	&	Std.Err.	&		&	Est.	&	Std.Err.	&		\\
\hline																				
Own National	&	($\tau=0$)	&	2.27E-05	&	2.12E-05	&		&	3.26E-05	&	2.05E-05	&		&	2.05E-05	&	1.55E-05	&		\\
Trad. Ads	&	($\tau=1$)	&	6.67E-05	&	2.15E-05	&	**	&	7.43E-05	&	2.08E-05	&	**	&	1.72E-05	&	1.57E-05	&		\\
	&	($\tau=2$)	&	2.66E-05	&	2.14E-05	&		&	6.41E-05	&	2.07E-05	&	**	&	3.98E-05	&	1.57E-05	&	*	\\
	&	($\tau=3$)	&	3.26E-05	&	2.14E-05	&		&	1.88E-05	&	2.07E-05	&		&	4.24E-05	&	1.56E-05	&	**	\\
	&	($\tau=4$)	&	6.99E-05	&	2.15E-05	&	**	&	3.22E-05	&	2.08E-05	&		&	3.94E-05	&	1.57E-05	&	*	\\
	&	($\tau=5$)	&	4.81E-05	&	2.11E-05	&	*	&	-1.15E-05	&	2.04E-05	&		&	9.08E-06	&	1.55E-05	&		\\
\end{tabular}}}
\end{center}
\end{table}

\renewcommand{\arraystretch}{1}
\begin{table}%[tb]
\caption{Parameter Estimates for Lagged D.V.'s in Descriptive Model}
%\vspace{1em}
\label{tb:lagdv1}
\begin{center}
{
\scriptsize{
\tabcolsep=0.15cm
\begin{tabular}{ll|ccl|ccl|ccl}
\hline																					
\multicolumn{2}{l|}{Brand Attitude D.V.}			&	\multicolumn{3}{c|}{Perceived Quality}					&	\multicolumn{3}{c|}{Perceived Value}					&	\multicolumn{3}{c}{Recent Satisfaction}					\\
\hline																					
\multicolumn{2}{l|}{Lagged Attitude}			&	Est.	&	Std.Err.	&		&	Est.	&	Std.Err.	&		&	Est.	&	Std.Err.	&		\\
\hline																					
Log Quality	&	($\tau=1$)	&	1.09E-01	&	2.66E-03	&	**	&	2.55E-02	&	2.52E-03	&	**	&	2.06E-02	&	1.91E-03	&	**	\\
	&	($\tau=2$)	&	4.33E-02	&	2.68E-03	&	**	&	1.90E-02	&	2.54E-03	&	**	&	1.64E-02	&	1.92E-03	&	**	\\
	&	($\tau=3$)	&	5.34E-02	&	2.68E-03	&	**	&	2.40E-02	&	2.54E-03	&	**	&	1.28E-02	&	1.92E-03	&	**	\\
	&	($\tau=4$)	&	3.97E-02	&	2.69E-03	&	**	&	1.16E-02	&	2.54E-03	&	**	&	1.11E-02	&	1.92E-03	&	**	\\
	&	($\tau=5$)	&	3.25E-02	&	2.69E-03	&	**	&	8.37E-03	&	2.55E-03	&	**	&	6.84E-03	&	1.93E-03	&	**	\\
	&	($\tau=6$)	&	3.70E-02	&	2.69E-03	&	**	&	1.29E-02	&	2.55E-03	&	**	&	1.74E-02	&	1.93E-03	&	**	\\
\end{tabular}}}
\end{center}
\end{table}

\renewcommand{\arraystretch}{1}
\begin{table}%[tb]
\caption{Parameter Estimates for Lagged D.V.'s in Model with Industry/Week Control}
%\vspace{1em}
\label{tb:lagdv2}
\begin{center}
{
\scriptsize{
\tabcolsep=0.15cm
\begin{tabular}{ll|ccl|ccl|ccl}
\hline																					
\multicolumn{2}{l|}{Brand Attitude D.V.}			&	\multicolumn{3}{c|}{Perceived Quality}					&	\multicolumn{3}{c|}{Perceived Value}					&	\multicolumn{3}{c}{Recent Satisfaction}					\\
\hline																					
\multicolumn{2}{l|}{Lagged Attitude}			&	Est.	&	Std.Err.	&		&	Est.	&	Std.Err.	&		&	Est.	&	Std.Err.	&		\\
\hline																						
Log Quality	&	($\tau=1$)	&	1.09E-01	&	2.75E-03	&	**	&	2.87E-02	&	2.65E-03	&	**	&	2.13E-02	&	2.01E-03	&	**	\\
	&	($\tau=2$)	&	4.76E-02	&	2.76E-03	&	**	&	2.32E-02	&	2.67E-03	&	**	&	1.46E-02	&	2.02E-03	&	**	\\
	&	($\tau=3$)	&	4.98E-02	&	2.77E-03	&	**	&	2.21E-02	&	2.67E-03	&	**	&	1.47E-02	&	2.03E-03	&	**	\\
	&	($\tau=4$)	&	3.81E-02	&	2.77E-03	&	**	&	1.30E-02	&	2.67E-03	&	**	&	1.03E-02	&	2.03E-03	&	**	\\
	&	($\tau=5$)	&	3.30E-02	&	2.77E-03	&	**	&	9.01E-03	&	2.68E-03	&	**	&	9.76E-03	&	2.03E-03	&	**	\\
	&	($\tau=6$)	&	3.79E-02	&	2.77E-03	&	**	&	1.47E-02	&	2.68E-03	&	**	&	1.57E-02	&	2.03E-03	&	**	\\
\end{tabular}}}
\end{center}
\end{table}

\renewcommand{\arraystretch}{1}
\begin{table}%[tb]
\caption{Parameter Estimates for Lagged D.V.'s in Model with Brand/Quarter Control}
%\vspace{1em}
\label{tb:lagdv3}
\begin{center}
{
\scriptsize{
\tabcolsep=0.15cm
\begin{tabular}{ll|ccl|ccl|ccl}
\hline																					
\multicolumn{2}{l|}{Brand Attitude D.V.}			&	\multicolumn{3}{c|}{Perceived Quality}					&	\multicolumn{3}{c|}{Perceived Value}					&	\multicolumn{3}{c}{Recent Satisfaction}					\\
\hline																					
\multicolumn{2}{l|}{Lagged Attitude}			&	Est.	&	Std.Err.	&		&	Est.	&	Std.Err.	&		&	Est.	&	Std.Err.	&		\\
\hline																						
Log Quality	&	($\tau=1$)	&	-6.35E-02	&	2.76E-03	&	**	&	2.16E-02	&	2.63E-03	&	**	&	1.47E-02	&	1.98E-03	&	**	\\
	&	($\tau=2$)	&	-1.16E-01	&	2.77E-03	&	**	&	1.88E-02	&	2.63E-03	&	**	&	1.29E-02	&	1.98E-03	&	**	\\
	&	($\tau=3$)	&	-1.01E-01	&	2.78E-03	&	**	&	2.66E-02	&	2.64E-03	&	**	&	1.11E-02	&	1.99E-03	&	**	\\
	&	($\tau=4$)	&	-1.08E-01	&	2.79E-03	&	**	&	1.70E-02	&	2.65E-03	&	**	&	1.12E-02	&	2.00E-03	&	**	\\
	&	($\tau=5$)	&	-1.09E-01	&	2.80E-03	&	**	&	1.58E-02	&	2.66E-03	&	**	&	7.85E-03	&	2.01E-03	&	**	\\
	&	($\tau=6$)	&	-9.82E-02	&	2.80E-03	&	**	&	2.02E-02	&	2.67E-03	&	**	&	1.84E-02	&	2.01E-03	&	**	\\
\end{tabular}}}
\end{center}
\end{table}

\renewcommand{\arraystretch}{1}
\begin{table}%[tb]
\caption{Parameter Estimates for Lagged D.V.'s in All Controls Model}
%\vspace{1em}
\label{tb:lagdv4}
\begin{center}
{
\scriptsize{
\tabcolsep=0.15cm
\begin{tabular}{ll|ccl|ccl|ccl}
\hline																					
\multicolumn{2}{l|}{Brand Attitude D.V.}			&	\multicolumn{3}{c|}{Perceived Quality}					&	\multicolumn{3}{c|}{Perceived Value}					&	\multicolumn{3}{c}{Recent Satisfaction}					\\
\hline																					
\multicolumn{2}{l|}{Lagged Attitude}			&	Est.	&	Std.Err.	&		&	Est.	&	Std.Err.	&		&	Est.	&	Std.Err.	&		\\
\hline																						
Log Quality	&	($\tau=1$)	&	-6.21E-02	&	2.86E-03	&	**	&	2.35E-02	&	2.77E-03	&	**	&	1.62E-02	&	2.09E-03	&	**	\\
	&	($\tau=2$)	&	-1.09E-01	&	2.87E-03	&	**	&	2.30E-02	&	2.77E-03	&	**	&	1.22E-02	&	2.10E-03	&	**	\\
	&	($\tau=3$)	&	-1.03E-01	&	2.88E-03	&	**	&	2.44E-02	&	2.79E-03	&	**	&	1.37E-02	&	2.11E-03	&	**	\\
	&	($\tau=4$)	&	-1.09E-01	&	2.89E-03	&	**	&	1.82E-02	&	2.80E-03	&	**	&	1.14E-02	&	2.11E-03	&	**	\\
	&	($\tau=5$)	&	-1.08E-01	&	2.90E-03	&	**	&	1.68E-02	&	2.81E-03	&	**	&	1.13E-02	&	2.12E-03	&	**	\\
	&	($\tau=6$)	&	-9.69E-02	&	2.91E-03	&	**	&	2.25E-02	&	2.82E-03	&	**	&	1.80E-02	&	2.13E-03	&	**	\\
\end{tabular}}}
\end{center}
\end{table}

\renewcommand{\arraystretch}{1}
\begin{table}%[tb]
\caption{Advertising Parameter Estimate Variation with $T_a$ in All-Controls Perceived Quality Model}
%\vspace{1em}
\label{tb:diff_lagsqua}
\begin{center}
{
\scriptsize{
\tabcolsep=0.05cm
\begin{tabular}{ll|rl|rl|rl|rl|rl|rl|rl}
\hline																															
\multicolumn{2}{l|}{Ad Par. Spec.}			&	\multicolumn{2}{c|}{1 Ad Lag}			&	\multicolumn{2}{c|}{2 Ad Lags}			&	\multicolumn{2}{c|}{3 Ad Lags}			&	\multicolumn{2}{c|}{4 Ad Lags}			&	\multicolumn{2}{c|}{5 Ad Lags}			&	\multicolumn{2}{c|}{6 Ad Lags}			&	\multicolumn{2}{c}{7 Ad Lags}			\\
\hline																															
Own	&	($\tau=0$)	&	2.33E-05	&		&	2.13E-05	&		&	2.38E-05	&		&	1.89E-05	&		&	2.27E-05	&		&	2.75E-05	&		&	3.47E-05	&		\\
National	&	($\tau=1$)	&	7.70E-05	&	**	&	6.89E-05	&	**	&	6.49E-05	&	**	&	7.07E-05	&	**	&	6.67E-05	&	**	&	6.94E-05	&	**	&	7.53E-05	&	**	\\
Trad.	&	($\tau=2$)	&		&		&	4.19E-05	&	*	&	2.87E-05	&		&	2.36E-05	&		&	2.66E-05	&		&	2.18E-05	&		&	2.49E-05	&		\\
Ads	&	($\tau=3$)	&		&		&		&		&	5.31E-05	&	*	&	3.58E-05	&		&	3.26E-05	&		&	3.66E-05	&		&	3.04E-05	&		\\
	&	($\tau=4$)	&		&		&		&		&		&		&	8.16E-05	&	**	&	6.99E-05	&	**	&	6.60E-05	&	**	&	7.20E-05	&	**	\\
	&	($\tau=5$)	&		&		&		&		&		&		&		&		&	4.81E-05	&	*	&	3.48E-05	&		&	2.98E-05	&		\\
	&	($\tau=6$)	&		&		&		&		&		&		&		&		&		&		&	5.68E-05	&	**	&	3.95E-05	&		\\
\end{tabular}}}
\end{center}
\end{table}

\renewcommand{\arraystretch}{1.5}
\begin{landscape}
\begin{table}%[tb]
\caption{Ad Parameter Estimates by Industry on Perceived Value}
%\vspace{1em}
\label{tb:effval_ind}
\begin{center}
{
\tiny{
\tabcolsep=0.01cm
\begin{tabular}{l|cccccc|cccccc|cccccc}
\hline																																					
	&	\multicolumn{6}{c|}{National Traditonal Ads}											&	\multicolumn{6}{c|}{Local Traditional Ads}											&	\multicolumn{6}{c}{Digital Ads}											\\
\hline																																					
	&	$\tau=0$	&	$\tau=1$	&	$\tau=2$	&	$\tau=3$	&	$\tau=4$	&	$\tau=5$	&	$\tau=0$	&	$\tau=1$	&	$\tau=2$	&	$\tau=3$	&	$\tau=4$	&	$\tau=5$	&	$\tau=0$	&	$\tau=1$	&	$\tau=2$	&	$\tau=3$	&	$\tau=4$	&	$\tau=5$	\\
\hline																																					
Consumer Goods	&	-2.15E-04	&	3.18E-04	&	-2.35E-04	&	2.59E-04	&	2.21E-05	&	-2.67E-05	&	2.05E-04	&	1.64E-04	&	2.86E-04	&	-2.41E-04	&	9.01E-05	&	1.16E-04	&	1.99E-05	&	-9.05E-05	&	-1.99E-04	&	1.28E-04	&	4.63E-05	&	-1.99E-04	\\
Tools/Hardware	&	-4.30E-04	&	-1.50E-05	&	-9.13E-05	&	-4.83E-04	&	-1.06E-04	&	-5.29E-05	&	-5.57E-04	&	2.51E-04	&	-3.53E-04	&	6.64E-04	&	6.74E-05	&	-4.46E-04	&	4.27E-04	&	-3.66E-04	&	-7.46E-04	&	3.67E-04	&	-1.16E-04	&	4.56E-04	\\
Soft Drinks	&	-2.64E-04	&	9.24E-06	&	-9.80E-05	&	1.30E-05	&	-1.27E-04	&	2.97E-04	&	6.82E-05	&	5.23E-05	&	3.67E-05	&	-1.53E-04	&	-2.62E-04	&	3.02E-05	&	5.61E-04	&	1.17E-04	&	3.92E-04	&	-4.70E-04	&	3.46E-04	&	3.31E-04	\\
Beverages: General	&	1.70E-05	&	-8.80E-06	&	1.45E-04	&	2.07E-04	&	4.78E-05	&	-8.22E-05	&	8.25E-05	&	-6.07E-05	&	-1.63E-04	&	-7.67E-05	&	6.30E-05	&	-4.45E-05	&	-3.14E-04	&	-8.65E-05	&	3.46E-04	&	3.07E-04	&	-1.71E-04	&	-3.96E-05	\\
Media Devices	&	-1.25E-04	&	4.59E-05	&	{\bf -2.79e-04}	&	-6.21E-05	&	5.04E-05	&	-1.96E-04	&	1.38E-05	&	7.29E-05	&	8.03E-05	&	1.22E-05	&	6.34E-05	&	1.22E-04	&	1.65E-04	&	-4.88E-04	&	4.41E-04	&	-2.01E-04	&	4.69E-04	&	-8.07E-05	\\
Drugs: OTC	&	1.50E-04	&	6.50E-06	&	-6.38E-05	&	-1.71E-05	&	-2.13E-04	&	2.16E-04	&	7.56E-05	&	2.47E-04	&	-2.04E-04	&	7.43E-05	&	8.74E-05	&	-1.52E-06	&	6.23E-05	&	1.04E-04	&	-9.16E-05	&	-2.33E-04	&	8.23E-06	&	3.34E-05	\\
\end{tabular}}}
\end{center}
\end{table}
\end{landscape}

\renewcommand{\arraystretch}{1.5}
\begin{landscape}
\begin{table}%[tb]
\caption{Ad Parameter Estimates by Industry on Recent Satisfaction}
%\vspace{1em}
\label{tb:effsat_ind}
\begin{center}
{
\tiny{
\tabcolsep=0.01cm
\begin{tabular}{l|cccccc|cccccc|cccccc}	
\hline																																					
	&	\multicolumn{6}{c|}{National Traditonal Ads}											&	\multicolumn{6}{c|}{Local Traditional Ads}											&	\multicolumn{6}{c}{Digital Ads}											\\
\hline																																					
	&	$\tau=0$	&	$\tau=1$	&	$\tau=2$	&	$\tau=3$	&	$\tau=4$	&	$\tau=5$	&	$\tau=0$	&	$\tau=1$	&	$\tau=2$	&	$\tau=3$	&	$\tau=4$	&	$\tau=5$	&	$\tau=0$	&	$\tau=1$	&	$\tau=2$	&	$\tau=3$	&	$\tau=4$	&	$\tau=5$	\\
\hline																																					
Consumer Goods	&	6.15E-05	&	1.49E-04	&	1.40E-04	&	1.10E-04	&	-1.49E-04	&	{\bf  4.61e-04}	&	-1.71E-04	&	1.52E-04	&	{\bf -3.56e-04}	&	1.67E-04	&	-7.75E-05	&	1.42E-04	&	2.08E-04	&	1.31E-04	&	-8.14E-06	&	1.08E-04	&	1.86E-04	&	-1.28E-04	\\
Tools/Hardware	&	4.53E-04	&	-1.13E-04	&	4.99E-05	&	{\bf  5.59e-04}	&	{\bf -6.70e-04}	&	-5.01E-05	&	4.72E-04	&	-2.61E-04	&	3.14E-04	&	-2.00E-04	&	1.73E-05	&	1.43E-05	&	-2.30E-04	&	1.64E-04	&	3.23E-04	&	3.11E-04	&	{\bf -1.47e-03}	&	{\bf  1.31e-03}	\\
Soft Drinks	&	-2.47E-04	&	1.56E-04	&	2.14E-05	&	7.94E-05	&	6.33E-05	&	1.99E-04	&	-1.08E-05	&	1.32E-04	&	1.19E-04	&	-1.06E-04	&	-1.22E-05	&	8.95E-05	&	-1.63E-05	&	-1.31E-04	&	3.41E-06	&	1.67E-04	&	-2.83E-04	&	1.58E-04	\\
Beverages: General	&	-7.95E-05	&	-1.53E-04	&	-1.63E-04	&	{\bf  1.92e-04}	&	1.48E-04	&	-7.35E-05	&	-9.86E-05	&	2.89E-05	&	-1.84E-04	&	-2.34E-05	&	4.89E-05	&	5.75E-05	&	2.47E-04	&	6.96E-05	&	-3.00E-05	&	-1.45E-04	&	8.36E-05	&	6.28E-05	\\
Media Devices	&	-7.55E-05	&	8.13E-06	&	-8.10E-05	&	1.39E-04	&	1.14E-04	&	6.45E-05	&	-1.27E-04	&	3.89E-05	&	-1.23E-04	&	-1.39E-05	&	-1.60E-06	&	3.58E-05	&	1.33E-04	&	-1.61E-04	&	2.58E-04	&	-1.45E-04	&	-2.08E-04	&	3.21E-06	\\
Drugs: OTC	&	6.70E-05	&	1.86E-04	&	1.29E-04	&	-5.67E-05	&	3.34E-05	&	8.78E-05	&	-1.14E-04	&	-8.50E-05	&	-3.02E-04	&	2.36E-05	&	-1.54E-04	&	6.46E-05	&	-3.54E-04	&	1.98E-04	&	1.08E-04	&	-2.42E-04	&	2.35E-04	&	-2.11E-04	\\
\end{tabular}}}
\end{center}
\end{table}
\end{landscape}

\renewcommand{\arraystretch}{1.3}
\begin{table}[h]
\caption{All-controls model estimated using 2-week data aggregation}
%\vspace{1em}
\label{tb:2wkpar}
\begin{center}
{
\footnotesize{
\tabcolsep=0.1cm
\begin{tabular}{ll|ccc|ccc|ccc}
\hline
\multicolumn{2}{l|}{Brand Attitude Models}      &       \multicolumn{3}{c|}{Perceived Quality}  &       \multicolumn{3}{c|}{Perceived Value}    &       \multicolumn{3}{c}{Recent Satisfaction} \\
\hline
&       &       Est.    &       Std.Err.        &       &       Est.    &       Std.Err.        &       &       Est.    &       Std.Err.        &       \\
\hline
Own Nat'l       &       ($\tau=0$)      &       6.86E-05        &       2.31E-05        &       **      &       8.29E-05        &       2.21E-05        &       **      &       2.03E-05        &       1.65E-05        &       \\
Trad. Ads       &       ($\tau=1$)      &       7.49E-05        &       2.30E-05        &       **      &       9.17E-05        &       2.20E-05        &       **      &       7.30E-05        &       1.64E-05        &       **      \\
&       ($\tau=2$)      &       9.76E-05        &       2.30E-05        &       **      &       5.07E-05        &       2.20E-05        &       *       &       4.47E-05        &       1.64E-05        &       **      \\
Own Loc.        &       ($\tau=0$)      &       4.03E-05        &       2.71E-05        &       &       6.20E-05        &       2.60E-05        &       *       &       -1.04E-05       &       1.93E-05        &       \\
Trad. Ads       &       ($\tau=1$)      &       7.63E-05        &       2.72E-05        &       **      &       3.11E-05        &       2.61E-05        &       &       -1.40E-05       &       1.94E-05        &       \\
&       ($\tau=2$)      &       3.55E-05        &       2.70E-05        &       &       2.45E-05        &       2.58E-05        &       &       -2.32E-06       &       1.92E-05        &       \\
\end{tabular}}}
\end{center}
\end{table}
\renewcommand{\arraystretch}{1.3}
\begin{table}[h]
\caption{All-controls model estimated using 4-week data aggregation}
%\vspace{1em}
\label{tb:4wkpar}
\begin{center}
{
\footnotesize{
\tabcolsep=0.1cm
\begin{tabular}{ll|ccc|ccc|ccc}
\hline
\multicolumn{2}{l|}{Brand Attitude Models}      &       \multicolumn{3}{c|}{Perceived Quality}  &       \multicolumn{3}{c|}{Perceived Value}    &       \multicolumn{3}{c}{Recent Satisfaction} \\
\hline
&       &       Est.    &       Std.Err.        &       &       Est.    &       Std.Err.        &       &       Est.    &       Std.Err.        &       \\
\hline
\multicolumn{2}{l|}{Own Nat'l Trad. Ads}        &       9.30E-05        &       2.94E-05        &       **      &       7.13E-05        &       2.80E-05        &       *       &       4.74E-05        &       2.06E-05        &       *       \\
\multicolumn{2}{l|}{Own Loc. Trad. Ads} &       1.72E-06        &       3.91E-05        &       &       2.42E-05        &       3.72E-05        &       &       -3.01E-05       &       2.73E-05        &       \\
\multicolumn{2}{l|}{Own Dig. Ads}       &       7.69E-05        &       3.85E-05        &       *       &       7.75E-06        &       3.66E-05        &       &       3.13E-05        &       2.69E-05        &       \\
\hline
\multicolumn{2}{l|}{Comp. Nat'l Trad. Ads}      &       -1.14E-03       &       5.56E-04        &       *       &       -1.07E-03       &       5.29E-04        &       *       &       2.51E-04        &       3.88E-04        &       \\
\multicolumn{2}{l|}{Comp. Loc. Trad. Ads}       &       -7.70E-04       &       4.38E-04        &       &       -5.10E-04       &       4.16E-04        &       &       -1.90E-04       &       3.06E-04        &       \\
\multicolumn{2}{l|}{Comp. Dig. Ads}     &       4.04E-04        &       4.61E-04        &       &       -1.09E-03       &       4.38E-04        &       *       &       1.70E-04        &       3.22E-04        &       \\
\end{tabular}}}
\end{center}
\end{table}
\end{document}